\documentclass[11pt]{article}
\usepackage{amsmath}
\usepackage{amsfonts}
\usepackage{amssymb}
\usepackage[pdftex]{color,graphicx}
\usepackage{authblk}

\usepackage{amsmath,amscd}

\newtheorem{Theorem}{Theorem}

\newtheorem{Cor}{Corollary}

\newtheorem{Lemma}{Lemma} 

\newtheorem{Def}{Definition}

\def\C{\mathbb{C}}

\newcommand{\ket}[1]{|{#1}\rangle}

\newcommand{\states}{\mathcal{S}}
\newcommand{\Aut}{\text{Sym}}

\title{Topological proofs of contextuality in quantum mechanics}
\author[1]{Cihan Okay}
\author[2]{Sam Roberts}
\author[2]{Stephen D. Bartlett}
\author[3]{Robert Raussendorf}
\affil[1]{Department of Mathematics, University of Western Ontario, London, Ontario, Canada}
\affil[2]{Centre for Engineered Quantum Systems, School of Physics, The University of Sydney, Sydney, NSW, Australia}
\affil[3]{Department of Physics and Astronomy, University of British Columbia, Vancouver, BC, Canada}
\date{\today}                     
\begin{document}

\maketitle

\begin{abstract}We provide a cohomological framework for contextuality of quantum mechanics that is suited to describing contextuality as a resource in measurement-based quantum computation. This framework applies to the parity proofs first discussed by Mermin, as well as a different type of contextuality proofs based on symmetry transformations. The topological arguments presented can be used in the state-dependent and the state-independent case. 
\end{abstract}

\section{Introduction}

Contextuality \cite{KS}-\cite{CSW} is a feature that distinguishes quantum mechanics from classical physics. To describe it, let's consider the question of whether it is possible to assign ``pre-existing'' outcomes to measurements of quantum observables which are merely revealed by measurement.  If this were possible, it would amount to a description of quantum mechanics in terms of classical statistical mechanics. Assuming such a model, for any two different sets ${\cal{A}}$ and ${\cal{B}}$ of mutually compatible observables containing a given observable $A$, it is reasonable to require that the value $\lambda(A)$ attached to the observable $A$ is a property of $A$ alone, and thus agrees in ${\cal{A}}$ and ${\cal{B}}$. ${\cal{A}}$ and ${\cal{B}}$ are measurement contexts for $A$, and the constraint on $\lambda(A)$ just described is called ``context independence''. Can context-independent pre-assigned outcomes $\lambda$, or probabilistic combinations thereof, describe all of quantum mechanics?---This turns out not to be the case \cite{KS}, \cite{Bell}, a fact which is often referred to as contextuality of quantum mechanics. 

For quantum computation, contextuality is a resource. In quantum computation with magic states \cite{BK} and in measurement-based quantum computation (MBQC) \cite{RB01}, no quantum speedup can occur without it \cite{How}--\cite{Qubit}; \cite{AB}--\cite{RR13}.

For the present work, the link between contextuality and quantum computation is the motivation to investigate the mathematical structures underlying contextuality. In this regard, 
Abramsky and coworkers have provided a sheaf-theoretic description of contextuality \cite{ABsheaf}. They have further identified cohomological obstructions to the existence of the  classical models described above, so-called non-contextual hidden variable models \cite{A2}, \cite{A3}. These methods, based on {\v{C}}ech cohomology,  have a wide range of applicability, covering the Bell inequalities \cite{Bell}, Hardy's model \cite{Hardy}, and the Greenberger-Horne-Zeillinger setting \cite{GHZ}.

Here, we provide a different cohomological framework for contextuality, involving group cohomology.  It is designed to describe the form of contextuality required for the functioning of measurement-based quantum computation. The connection between contextuality and MBQC was first observed in the example of Mermin's star \cite{AB}, and subsequently extended to all MBQC on multi-qubit states \cite{Hob}, \cite{RR13}.
From the latter works it is known that all contextuality proofs relevant for MBQC are generalizations of Mermin's star, in the sense that they invoke an algebraic contradiction to the existence of even a single non-contextual consistent value assignment. By its intended scope, the present framework only needs to apply to such kinds of proofs. But then there is an additional requirement: the cohomological framework in question needs to reproduce the original parity proofs in a topological guise. The reason for this requirement is that both the parity proofs and the classical side-processing required in every MBQC are based on {\em{the same}} linear relations (See Appendix~\ref{ContextMBQC} for a summary on contextuality in MBQC; also see \cite{RR16}).

Next to the parity-based proofs of contextuality exemplified by Mermin's square and star, we investigate a different type of contextuality proof which is  based on symmetry. The central object in these proofs is the group of transformations that leave the set of observables involved in a parity-based contextuality proof invariant, up to phases.  We show that nontrivial cohomology of the symmetry group implies contextuality. Furthermore, the parity-based and the symmetry-based contextuality proofs are related. Every symmetry-based proof implies a parity-based proof.
 
To summarize, we examine proofs of contextuality of quantum mechanics that have two attributes. They can either be parity-based or symmetry-based, and be state-independent or state-dependent. There are thus four combinations, and for each of these types of proofs we present a topological formulation. The parity-based contextuality proofs are discussed in Section~\ref{PbP} and the symmetry-based  proofs in Section~\ref{SbP}.

\section{First example}\label{Msqr}

\begin{figure}
\begin{center}
\begin{tabular}{lcl}
(a) && (b)\\
\includegraphics[height=4.5cm]{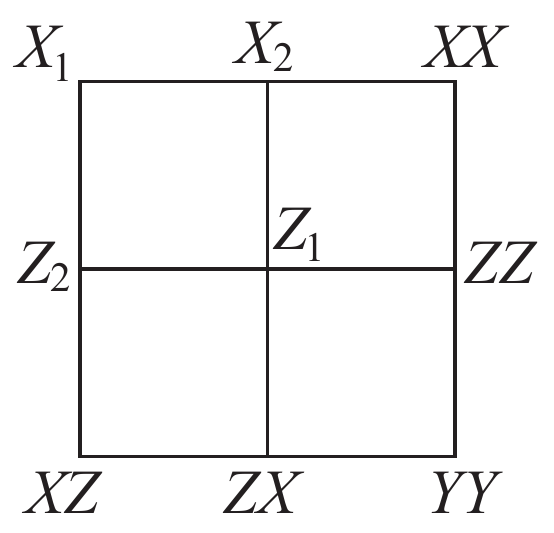} &&
\includegraphics[height=4.5cm]{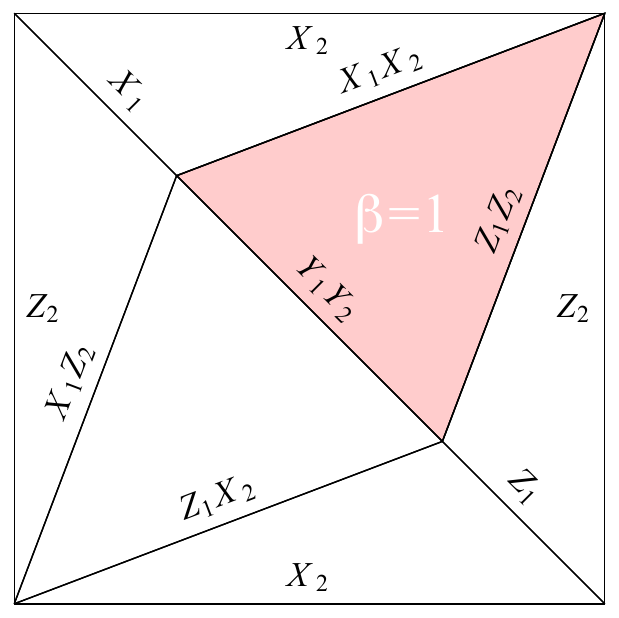}
\end{tabular}
\caption{\label{MermSq} Mermin's square \cite{Merm}. (a) Each horizontal and vertical line corresponds to a measurement context. Each context is composed out of three commuting Pauli observables $A, B, C$ which satisfy the constraint $ABC=\pm I$. (b) Mermin's square re-arranged on a surface. The 9 Pauli observables are now associated with the edges, and each measurement context is associated with the boundary of one of the six elementary faces. The exterior edges are identified as shown.}
\end{center}
\end{figure}

To illustrate what ``reproducing the original parity proofs in topological guise'' means, we consider as a first example Mermin's square \cite{Merm} (also see \cite{Peres}), one of the simplest proofs of contextuality of quantum mechanics. Mermin's square, depicted in Fig.~\ref{MermSq}a, demonstrates that in Hilbert spaces of dimension $\geq 4$ it is impossible to consistently assign pre-existing values to all quantum mechanical observables.

Each row and each column of the square represents a measurement context, consisting of commuting observables. Furthermore, the observables in each context multiply to $\pm I$. For example, in the bottom row in Fig.~\ref{MermSq}a, we have $(X_1Z_2)(Z_1X_2)(Y_1Y_2)=+I$, and in the right column $(X_1X_2)(Z_1Z_2)(Y_1Y_2)=-I$.  Now assume the nine Pauli observables $T_a$ in the square have pre-existing context-independent outcomes $\lambda(T_a)=(-1)^{s(T_a)}$, with $s(T_a)\in \mathbb{Z}_2$ (the eigenvalues of the Pauli observables are $\pm 1$). Then, the product relations among the observables translate into constraints among the consistent value assignments. Continuing with the above-stated relations, we obtain the constraints $\lambda(X_1Z_2)\lambda(Z_1X_2)\lambda(Y_1Y_2)=1$, and $\lambda(X_1X_2)\lambda(Z_1Z_2)\lambda(Y_1Y_2)=-1$. It is convenient to express these relations in terms of the value assignments $s(\cdot)$ rather than the measured eigenvalues $\lambda(\cdot)$. This leads to a system of linear equations,
\begin{equation}\label{Mcons}
\begin{array}{rcl}
s(X_1) + s(X_2) +s(X_1X_2) \mod 2 &=& 0,\\
s(Z_2) + s(Z_1) + s(Z_1Z_2) \mod 2 &=& 0,\\
s(X_1Z_2)+ s(Z_1X_2)+ (Y_1Y_2)\mod 2 &=& 0,\\
s(X_1)+ s(Z_2) +s(X_1Z_2) \mod 2 &=& 0,\\
s(Z_1) + (X_2)+ s(Z_1X_2) \mod 2&=& 0,\\
s(X_1X_2) + s(Z_1Z_2) + s(Y_1Y_2) \mod 2 &=& 1.
\end{array}
\end{equation}
No assignment $s$ can satisfy these relations. To see this, add the above equations mod 2, and observe that each value $s(T_a)$ appears twice on the left hand side. This results in the contradiction 0 = 1. 

We now reproduce this contradiction in a topological fashion. For this purpose, the six observables are regarded as labeling the edges in a tessellation of a torus; See Fig.~\ref{MermSq}b. The value assignment $s$ is now a 1-cochain. Denote by $f$ any of the six elementary faces of the surface, such that $\partial f =a+b+c$, for three edges $a$, $b$, $c$. Then there is a binary-valued function $\beta$ defined on the faces $f$ such that $T_c= (-1)^{\beta(a,b)} T_aT_b$. As before, these product constraints among (commuting) observables induce constraints among the corresponding values, namely $s(a)+s(b)+s(c) \mod 2 =\beta(f)$. By dialing through the six faces $f$, we reproduce the six constraints of Eq.~(\ref{Mcons}).

These constraints have a topological interpretation. Namely, $\beta$ is a 2-cochain, and, for any consistent context-independent value assignment $s$ it holds that
\begin{equation}
ds = \beta.
\end{equation}
Therein, $d$ the coboundary operator and the addition is $\text{mod}\; 2$. We can now show that for the present function $\beta$, which evaluates to 0 on 5 faces and to 1 on one face, no consistent value assignment $s$ exists. To this end, we integrate over the whole surface $F$ which is a 2-cycle, $\partial F = 0$. By Stokes' theorem,
$$
1 = \int_F \beta  =\int_F ds = \oint_{\partial F} s =\oint_0 s =0,
$$
where all integration is mod 2. In chain/cochain notation, this reads $1=\beta(F)=ds(F)=s(\partial F)=s(0)=0$. This is the same contradiction as above in Eq.~(\ref{Mcons}), but in cohomological form.  As we show in Section~\ref{PbP} of this paper, all parity proofs consisting of a set of conflicting linear constraints of the form Eq.~(\ref{Mcons}) can be given a similar cohomological interpretation.

To conclude this section, we remark that the above topological version of Mermin's square, in its mathematical structure, resembles a certain aspect of electromagnetism \cite{GKEM}. First, consider the vector calculus question of whether a given vector field $\textbf{B}$ can be written as the curl of some vector potential $\textbf{A}$, i.e., $\textbf{B} = \nabla \times \textbf{A}$. This possibility is ruled out by the existence of a closed surface $F$ for which $\int_F d\textbf{F}\cdot \textbf{B} \neq 0$. Here, $\textbf{A}$ is a 1-cochain (1-form) and $\textbf{B}$ is a 2-cochain (2-form). They are the counterparts of the value assignment $s$ and the function $\beta$, respectively. Now let $\textbf{B}$ be a magnetic field. The statement $\int_F d\textbf{F}\cdot \textbf{B} \neq 0$ for some closed surface $F$---the counterpart of a contextuality proof $\beta(F) \neq 0$---would indicate the presence of magnetic monopoles. However, in contrast to contextuality \cite{SqTest}, magnetic monopoles---while being a theoretical possibility---have to date not been experimentally observed \cite{Polch}.

\section{Measurement and contextuality}

In this section we define our measurement setting and notion of contextuality.

\subsection{Observables}\label{PC}

In this paper, we consider observables with a restriction on their eigenvalues. Specifically, the eigenvalues are all of the form $\omega^k$, where $\omega = e^{2\pi i/d}$, for some $d \in \mathbb{Z}$, and $k \in \mathbb{Z}_d$.  For $d>2$, such observables are in general not Hermitian operators. However, that doesn't matter. We may look at the measurement of these observables in two equivalent ways. (i) The observables are unitary, and their eigenvalues can thus be found by phase estimation. Further, due to the special form of the eigenvalues, phase estimation is exact. (ii) If $O=\sum_i \omega^{s_i} |i\rangle\langle i|$, with all $s_i \in \mathbb{Z}_d$, one may instead measure $\tilde{O}= \sum_i s_i |i\rangle\langle i|$, which is Hermitian and has the same eigenspaces as $O$.---We note that non-Hermitian observables have found use in Bell inequalities with more than two outcomes per party \cite{Zycz}, and also in contextuality proofs \cite{Zuk1}, \cite{Zuk2}. \medskip

Out of the set of observables $\cal{O}$, we identify an indexed set $\{ T_a, a\in E\}$ over a set $E$.  Every observable $O \in \cal{O}$ is related to an element $T_a$ from this indexed set by a phase $\omega^k$ for some $k$.  That is, $\cal{O}$  is of the form 
\begin{equation}\label{ObsSet}
{\cal{O}}= \{\omega^{k}T_a|\, a\in E, k\in \mathbb{Z}_d\}.
\end{equation}
For example $\cal{O}$ can be taken to be all of the Pauli observables and $E$ corresponds to the set of Pauli observables up to a phase. The set $E$ has more structure which comes from the multiplicative structure of  $\cal{O}$: We require that the product of two operators $T_a$ and $T_b$ belongs to ${\cal O}$ if they commute, $[T_a,T_b]=0$. For commuting operators the product $T_aT_b$ will correspond to an operator $T_c$ up to a phase. We write $c=a+b$ for this unique element in $E$.
The operators $\{T_a\}_{a\in E}$ satisfy the relation 
\begin{equation}\label{3T}
T_{a+b} = \omega^{\beta(a,b)} T_aT_b,\;\; \forall a,b \in E,\,\text{s.th.}\, [T_a,T_b]=0.
\end{equation}
The function $\beta$ takes values in $\mathbb{Z}_d$. To see this, consider the simultaneous eigenvalues of the operators $T_a$, $T_b$, $T_{a+b}$. With Eq.~(\ref{3T}) it holds that $\omega^{k_{a+b}}=\omega^{\beta(a,b)+k_a+k_b}$, and $k_{a+b}, k_a, k_b\in \mathbb{Z}_d$. Thus $\beta(a,b)\in \mathbb{Z}_d$, as stated.

For any triple $\{T_a,T_b,T_{a+b}\}$ of observables satisfying the commutativity condition $[T_a,T_b]=0$, the simultaneous eigenvalues can be measured. While individually random, the measurement outcomes are strictly correlated, $\lambda(a+b)/\lambda(a)\lambda(b) =\omega^{\beta(a,b)}$. These correlations, which are predicted by quantum mechanics and are verifiable by experiment, form the basis of Mermin's state-independent contextuality proofs \cite{Merm}.  The function $\beta$ is thus a central object in present discussion, summing up the physical properties of ${\cal{O}}$.

\subsection{Definition of contextuality}

We now define the notion of a non-contextual hidden-variable model (ncHVM) with definite value assignments. First, a measurement context is a commuting set $M \subset {\cal{O}}$. The set of all measurement contexts is denoted by ${\cal{M}}$.

\begin{Def}\label{HVM1} Consider a quantum state $\rho$ and a set ${\cal{O}}$ of observables grouping into contexts $M \in {\cal{M}}$ of simultaneously measurable observables.
 A non-contextual hidden variable model $(\states, q_\rho,\Lambda)$  consists of a probability distributon $q_\rho$ over a set $\states$ of internal states and a set $\Lambda=\{\lambda_\nu\}_{\nu\in\states}$ of value assignment functions  $\lambda_\nu: {\cal{O}}\rightarrow\C$ that meet the following criteria.
 \begin{itemize}
  \item[(i)]  Each $\lambda_\nu\in\Lambda$ is consistent with quantum mechanics: for any set $M \in{\cal{O}}$ of commuting observables there exists a quantum state $\ket{\psi}$ such that
  \begin{equation}\label{eq:JointEigenvalue}
  A\ket{\psi}=\lambda_\nu(A)\ket{\psi}, \quad\forall A\in M.
  \end{equation}
   \item[(ii)] The distribution $q_\rho$ satisfies
   \begin{equation}
   \mathrm{tr}(A \rho)=\sum_{\nu\in\mathcal{S}} \lambda_\nu(A)q_\rho(\nu), \quad \forall A\in {\cal{O}}
   \end{equation}
  \end{itemize}
 \end{Def}
Condition (i) in Definition~\ref{HVM1} means that for every internal state $\nu$ of the non-contextual HVM the corresponding value assignment $\lambda_\nu$ is consistent across measurement contexts. \smallskip
 
We say that a physical setting $(\rho,{\cal{O}})$ is contextual if it cannot be described by any ncHVM $(\states, q_\rho,\Lambda)$.

\begin{Lemma}\label{ValAssConstr}
For any triple $A,B,AB \in {\cal{O}}$ of simultaneously measurable observables and any internal state $\nu \in {\cal{S}}$ of an ncHVM   $(\states, q_\rho,\Lambda)$ it holds that
\begin{equation}\label{Cons}
\lambda_\nu(AB) =  \lambda_\nu(A)  \lambda_\nu(B).
\end{equation}
\end{Lemma}
The relation Eq.~(\ref{Cons}) was first used in \cite{Merm} to rule out the existence of deterministic value assignments for Mermin's square and star. In the same capacity it is also used in the present discussion.\medskip  

{\em{Proof of Lemma~\ref{ValAssConstr}.}} Consider a set $M=\{A,B,AB\} \subset {\cal{O}}$ of observables such that $[A,B]=0$. This set qualifies as a possible $M$ in the sense of point (i) of Def.~\ref{HVM1}. Therefore, for any $\nu \in \Lambda$ there exists a quantum state $|\psi\rangle$ such that 
$
A |\psi\rangle = \lambda_\nu(A)\,|\psi\rangle$, $B |\psi\rangle = \lambda_\nu(B)\,|\psi\rangle$, $AB |\psi\rangle = \lambda_\nu(AB)\,|\psi\rangle.
$
Furthermore, $(AB)|\psi\rangle=A(B|\psi\rangle)=\lambda_\nu(A)\lambda_\nu(B)|\psi\rangle$. By comparison, $ \lambda_\nu(AB) =  \lambda_\nu(A)  \lambda_\nu(B)$, which proves Eq.~(\ref{Cons}). $\Box$

\section{Parity-based contextuality proofs}\label{PbP}

The example of Section~\ref{Msqr} is not special. As we show here, every parity-based contextuality proof---consisting of a set of conflicting linear constraints on the value assignments as in Eq.~(\ref{Mcons})---can be given a cohomological formulation. The main result of this section is Theorem~\ref{T1}.

\subsection{The chain complex ${\cal{C}}_*$}\label{Complex}

We have two assumptions on the set of operators $\cal{O}$:
\begin{enumerate}
\item $\cal{O}$ is closed under products of commuting operators i.e., if $[O_1,O_2]=0$ for $O_1,O_2\in {\cal{O}}$ then $O_1O_2\in \cal{O}$.
\item $\cal{O}$ contains the identity operator.
\end{enumerate}
 
Let $\eta:E\rightarrow {\cal O}$ denote the map given by
\begin{equation}
\label{etaDef}
\eta(a)=T_a,
\end{equation}
with $E$ the index set introduced in Eq.~(\ref{ObsSet}).
The set $E$ has more structure coming from Eq.~(\ref{3T}).  We say two elements $a,b\in E$ \textit{commute} if the corresponding operators commute $[T_a,T_b]=0$.
Given two commuting elements $a,b\in E$  we define the sum $a+b\in E$ to be the unique element which satisfies 
$T_{a+b}=\omega^{\beta(a,b)}T_aT_b$, cf. Eq.~(\ref{3T}). We assume that there is an element in $E$ denoted by $0$ corresponding to the identity operator $\eta(0)=I$ in $\cal{O}$. Under this addition operation every maximal subset of commuting elements in $E$  has the structure of an abelian group.   

Let us define the chain complex ${\cal{C}}_*={\cal{C}}_*(E)$. A standard reference for chain complexes  is \cite{Weibel}.
It will suffice to describe this complex up to dimension three, i.e., ${\cal{C}}_*= \{C_0, C_1,C_2,C_3\}$. The geometric picture is as follows. The space we consider consists of a single vertex ($0$-cell). It has an edge ($1$-cell) for each element of the set $E$ whose both boundary points attached to the single vertex. A face ($2$--cell) is attached for every product relation among commuting operators. The set of faces is thus given by 
\begin{equation}\label{faces}
F=\{ (a,b)\in E\times E|\; [T_a,T_b]=0 \}.
\end{equation}
Thus, every face $(a,b)\in F$ is bounded by three edges, namely $a$, $b$ and $a+b$.

\begin{figure}
\begin{center}
\parbox{6cm}{\includegraphics[width=6cm]{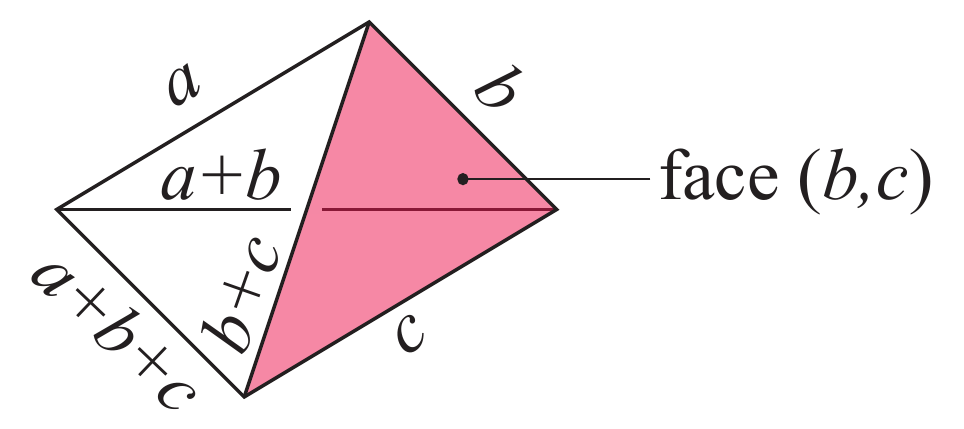}}
\caption{\label{VolDef} An elementary volume $V \in C_3$, bounded by four faces.} 
\end{center}
\end{figure}

Volumes ($3$-cells) are constructed from triples of commuting observables $T_a,T_b,T_c$ (see Fig.~\ref{VolDef} for an illustration). The set of volumes is
\begin{equation}\label{Volumes}
V=\{(a,b,c)\in E\times E\times E|\;[T_a,T_b]=[T_b,T_c]=[T_a,T_c]=0\}.
\end{equation}
Now comes the description of the chains:
\begin{enumerate}
\item $C_0=\mathbb{Z}_d$ since there is a single vertex.
\item $C_1=\mathbb{Z}_d E$, i.e., the elements of $C_1$ are linear combinations
$$
\sum_{a\in E} \alpha_a [a]\;\; \text{where $\alpha_a\in \mathbb{Z}_d$.}
$$
In other words, $C_1$ is freely generated as a $\mathbb{Z}_d$-module by  $[a]$, where $a\in E$.
\item $C_2$ is freely generated as a $\mathbb{Z}_d$-module by the pairs $[a|b]$, where $(a,b)\in F$.
\item $C_3$ is freely generated as a $\mathbb{Z}_d$-module by the triples $[a|b|c]$, where $(a,b,c)\in V$.
\end{enumerate} 
In summary $C_1,C_2,C_3$ are freely generated by $E,F,V$ as $\mathbb{Z}_d$-modules.
We stop at  dimension three although the definition can be continued for higher dimensions analogously, see \cite{A-C-T}. The differentials in the complex
$$
C_3\stackrel{\partial}{\rightarrow} C_2 \stackrel{\partial}{\rightarrow} C_1 \stackrel{\partial}{\rightarrow} C_0
$$
are defined by
$$
\partial[a]=0,\;\; \partial[a|b]=[b]-[a+b]+[a],\;\; \partial[a|b|c]=[b|c]-[a+b|c]+[a|b+c]-[a|b].
$$
Here the general pattern  is as follows
$$
\partial [a_1|a_2|\cdots|a_n] = [a_2|a_3|\cdots|a_n] + \sum_{i=1}^{n-1} (-1)^i [a_1|\cdots|a_i+a_{i+1}|\cdots|a_n] +(-1)^n [a_1|a_2|\cdots|a_{n-1}].
$$
The homology groups of ${\cal{C}}_*$ are defined by 
$$
H_n(\mathcal{C}_*,\mathbb{Z}_d)=\frac{\text{ker}(\partial)}{\text{im}(\partial)}.
$$
The dual notion of cochains ${\cal C}^*$ gives a cochain complex
$$
C^3\stackrel{d}{\leftarrow} C^2 \stackrel{d}{\leftarrow} C^1 \stackrel{d}{\leftarrow} C^0
$$
where $C^n$ consists of $\mathbb{Z}_d$-module maps $\phi:C_n\rightarrow \mathbb{Z}_d$.
The differential $d: C^n \rightarrow C^{n+1}$ is defined by
$
d\phi(\alpha)=\phi(\partial\alpha)
$
where $\alpha\in C_{n+1}$.  

\subsection{$\beta$ is a 2-cocycle}\label{beta}

We may now formally extend the function $\beta$ introduced in Eq.~(\ref{3T}) from $F$ to all of $C_2$
via the linear relations $\beta(u+v) = \beta(u)+\beta(v)$, $\beta(ku)=k\beta(u)$, for all $u,v\in C_2$, $k\in \mathbb{Z}_d$. The function $\beta$ is thus a 2-cochain, $\beta\in C^2$.

The function $\beta$ is constrained in the following way. Consider three commuting elements $a,b,c \in E$, and expand the observable $T_{a+b+c}$ in two ways,
$$
\begin{array}{rcl}
T_{a+b+c} &=& \displaystyle{T_{(a+b)+c}}\\
&=& \displaystyle{\omega^{\beta(a+b,c)}T_{a+b}T_c}\\
&=&\displaystyle{\omega^{\beta(a+b,c)+\beta(a,b)}T_aT_bT_c},
\end{array}
$$
and
$$
\begin{array}{rcl}
T_{a+b+c} &=& \displaystyle{T_{a+(b+c)}}\\
&=& \displaystyle{\omega^{\beta(a,b+c)}T_aT_{b+c}}\\
&=& \displaystyle{\omega^{\beta(a,b+c)+\beta(b,c)}T_aT_bT_c}.
\end{array}
$$
Comparing the two expressions, we find that
\begin{equation}\label{betaConstr}
\beta(a+b,c)+\beta(a,b)-\beta(a,b+c)-\beta(b,c) \mod d = 0,
\end{equation}
whenever  $[T_a,T_b]=0$, $[T_a,T_c]=0$, and $[T_b,T_c]=0$. 

The four faces $(a,b)$, $(a+b,c)$, $(a,b+c)$, $(b,c)$, with appropriate orientation (hence sign), bound a volume $V$, i.e.,
$$
\partial V = (a+b,c)+(a,b)-(a,b+c)-(b,c).
$$
Geometrically, the situation looks as displayed in Fig.~\ref{VolDef}. We can follow the convention that $(a,b)$ denotes a face in the geometric sense and $[a|b]$ denotes an element of the chain complex. So $\partial V =[a+b|c]+[a|b]-[a|b+c]-[b|c]$.
Therefore, with Eq.~(\ref{betaConstr}), 
$$
\begin{array}{rcl}
d\beta (V) = \beta(\partial V) &=& \beta ((a+b,c)+(a,b)-(a,b+c)-(b,c))\\
&=& \beta(a+b,c)+\beta(a,b)-\beta(a,b+c)-\beta(b,c)\\
&=& 0.
\end{array}
$$
Applying this relation to all volumes $V\in C_3$, we obtain
\begin{equation}\label{BetaConstr}
d\beta \equiv 0.
\end{equation}
Finally, there is an equivalence relation among the functions $\beta$.
To see this, recall the map $\eta: E \longrightarrow {\cal{O}}$ which is defined by $a\mapsto T_a$. There is a certain freedom in this definition which does not affect the commutation relations of the operators. Consider the following re-parametrization
\begin{equation}\label{EtaTrans}
\eta_{\gamma}(\cdot)=\omega^{\gamma (\cdot)} \eta(\cdot),
\end{equation}
where $\gamma: E \longrightarrow \mathbb{Z}_d$. Then $[\eta(a),\eta(b)]=0$ if and only if $[\eta_\gamma(a),\eta_\gamma(b)]=0$.
From the perspective of contextuality, it does not matter which map $\eta_{\gamma}$ we use to define the observables $\{T_a, a \in E\}$. Contextuality cannot be defined away by rephasing. However, the function $\beta$ is affected by the transformation Eq.~(\ref{EtaTrans}). Namely, changing from $\eta_0=\eta$ to $\eta_\gamma$ results in 
\begin{equation}\label{BetaIdent}
\begin{array}{rcl}
\beta(a,b) \longrightarrow \beta_\gamma(a,b) &=&\beta(a,b) -\gamma(a)-\gamma(b)+\gamma(a+b)\\
&=&\beta(a,b) - d\gamma(a,b).
\end{array}
\end{equation}
Therein, all addition is $\text{mod}\; d$. The functions $\beta$ are thus subject to a restriction Eq.~(\ref{BetaConstr}) and an identification Eq.~(\ref{BetaIdent}). The various possible functions $\beta$ thus fall into equivalence classes $[\beta]=\{\beta + d\gamma,\forall \gamma\}$, and hence $[\beta]\in H^2({\cal{C}},\mathbb{Z}_d)$.

\subsection{Cohomological formulation of parity-based contextuality proofs}\label{CohPar}

The function $\beta$ relates to the question of existence of non-contextual HVMs.  We have the following result.
First, a non-contextual value assignment $s: E \longrightarrow \mathbb{Z}_d$, is such that $\lambda(T_a) = \omega^{s(a)}$. Again, by linearity, we can extend the assignment from $E$ to all of $C^1$, and $s$ is thus a 1-cochain. We have the following relation.
\begin{Lemma}\label{ConsistBeta}
For every consistent non-contextual value assignment $s: E \longrightarrow \mathbb{Z}_d$ it holds that
\begin{equation}\label{ValTest}
ds = -\beta.
\end{equation}
\end{Lemma}
{\em{Proof of Lemma~\ref{ConsistBeta}.}} Evaluating Eq.~(\ref{ValTest}) on any given face $(a,b)\in F$ reads
\begin{equation}\label{ValFace}
s(a)+s(b)-s(a+b) =-\beta(a,b).
\end{equation}
As a consequence of Eq.~(\ref{Cons}), $\lambda(\omega^x A) =\omega^x \lambda(A)$, for all $x\in \mathbb{Z}_d$ and all $A \in {\cal{O}}$. Now, with Lemma~\ref{ValAssConstr}, setting $A=T_a$ and $B=T_b$ in Eq.~(\ref{Cons}), it holds that $\lambda(T_a)\lambda(T_b) =\lambda(T_aT_b)=\lambda(\omega^{-\beta(a,b)}T_{a+b}) = \omega^{-\beta(a,b)} \lambda(T_{a+b})$. This is precisely what Eq.~(\ref{ValFace}) requires. $\Box$
\medskip

\begin{Theorem}\label{T1}
Given set ${\cal{O}}$ of observables, if $H^2({\cal{C}},\mathbb{Z}_d) \ni [\beta]\neq 0$ then ${\cal{O}}$ exhibits state-independent contextuality.
\end{Theorem}

{\em{Proof of Theorem~\ref{T1}.}} If there were a value assignment $s$ it would satisfy $ds=-\beta$. This means that $\beta$ is a boundary: $\beta=d(-s)$. Hence $[\beta]=0$. $\Box$\medskip

{\em{Example: Mermin's star.}} In addition to Mermin's square, which we already discussed in Section~\ref{Msqr}, we now provide Mermin's star \cite{Merm} as a further example. Mermin's star comes both in a state-independent and a state-dependent version, and is thus best suited as a running example for all topological constructions presented in this paper.

\begin{figure}
\begin{center}
\begin{tabular}{lclcl}
(a) &&(b) && (c)\\
\includegraphics[height=4.4cm]{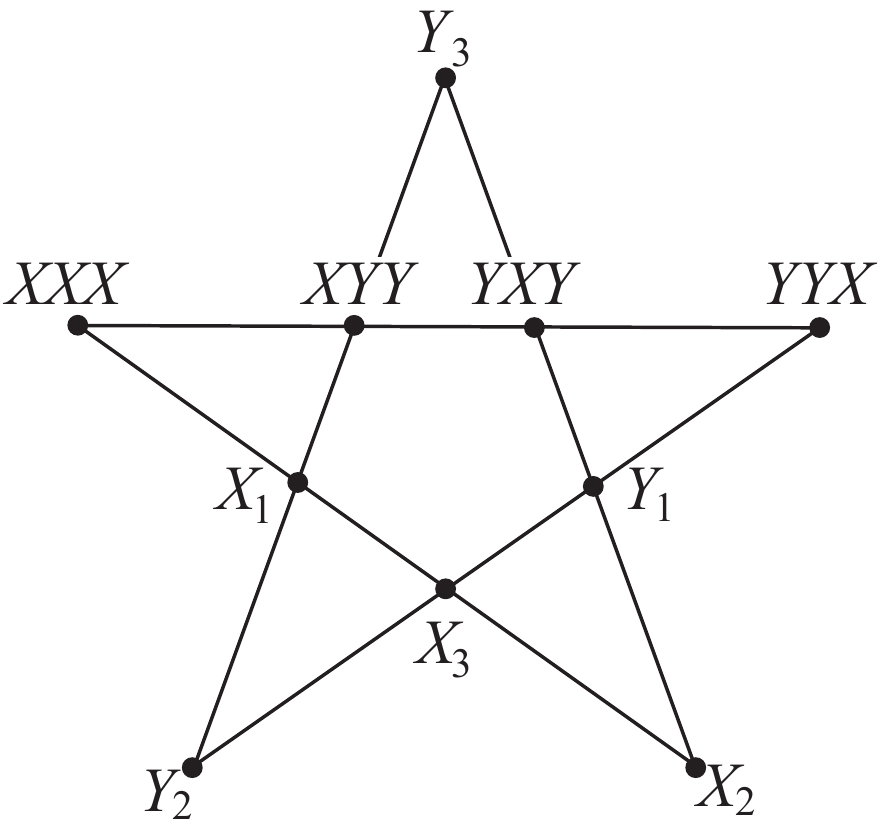} && \includegraphics[height=4.4cm]{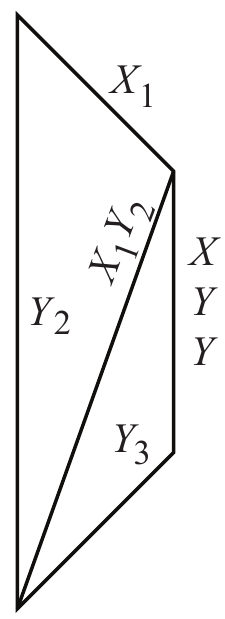} &&  \includegraphics[height=4.4cm]{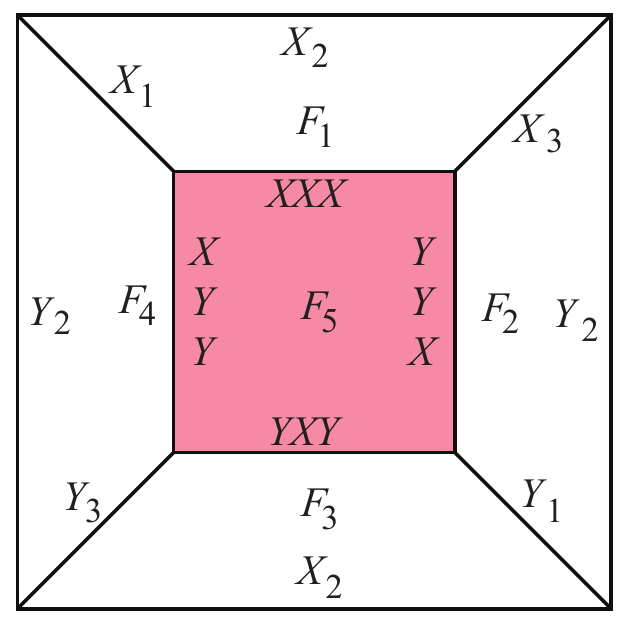}
\end{tabular}
\caption{\label{MS1} (a) The state-independent version of Mermin's star \cite{Merm}. (b) Two elementary three-sided faces combining to a four-sided surface. (c) Topological representation of Mermin's star. The left and right edges and the top and bottom edges, respectively, are identified.  }
\end{center}
\end{figure}

Here we consider the state-independent version; See Fig.~\ref{MS1}. Denote by $F_\text{star}$ the surface displayed in Fig.~\ref{MS1}c, consisting of the five smaller surfaces $F_1$,..,$F_5$ each corresponding to a measurement context in Fig.~\ref{MS1}a. Each of the surfaces $F_i$ may be split up into two elementary faces; See Fig.~\ref{MS1}b. $F_\text{star}:=\sum_{i=1}^5 F_i$ satisfies $\partial F_\text{star}=0$. Since $(X_1X_2X_3)(X_1Y_2Y_3)(Y_1X_2Y_3)(Y_1Y_2X_3)=-I$, we have $\beta(F_5)=1$, and for the other four measurement contexts it holds that $\beta(F_i)=0$. Hence, $\beta(F_\text{star})=1$. If $\beta=ds$ for some 1-cochain $s$, then $1=\beta(F_\text{star})=ds(F_\text{star})=s(\partial F_\text{star})=s(0)=0$. Contradiction. Hence, $[\beta]\neq 0$. Then, by Theorem~\ref{T1}, Mermin's star exhibits state-independent contextuality, in accordance with the original proof \cite{Merm}. 

\subsection{Squaring the star}

It tuns out that, from the cohomological perspective developed above, Mermin's square and star are {\em{equivalent}} contextuality proofs. Denote by ${\cal{C}}_*(3)$ the complex induced by the set ${\cal{O}}=\mathbb{P}^3$, the Pauli observables on 3 qubits. Both Mermin's square and star embed into it. The star provides a closed surface $F_\text{star} \in C_2(3)$ and the square provides a closed surface $F_\text{square}\in C_2(3)$, such that  $\beta(F_\text{star})=1$ and $\beta(F_\text{square})=1$. Both facts thus equally demonstrate that $\beta\neq 0 \in H^2({\cal{C}}_*(3),\mathbb{Z}_2)$.

What makes the star and the square equivalent is that there is a volume $V \in C_3(3)$ such that
\begin{equation}\label{SurfEq}
F_\text{square} = F_\text{star} + \partial V.
\end{equation}
The surfaces $F_\text{square}$ and $F_\text{star}$ representing the respective contextuality proofs are elements of the {\em{same}} homology class in $H_2({\cal{C}}_*(3),\mathbb{Z}_2)$; and therefore $\beta(F_\text{square})=\beta(F_\text{star})$ for any 2-cocycle $\beta$.

\begin{figure}
\begin{center}
\begin{tabular}{lcl}
(a) && (b)\\
\includegraphics[height=4.7cm]{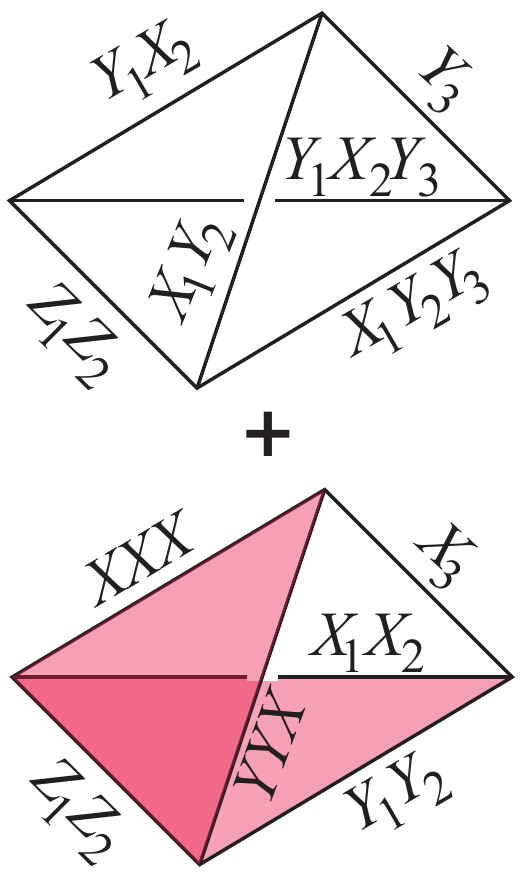} && \includegraphics[height=4.7cm]{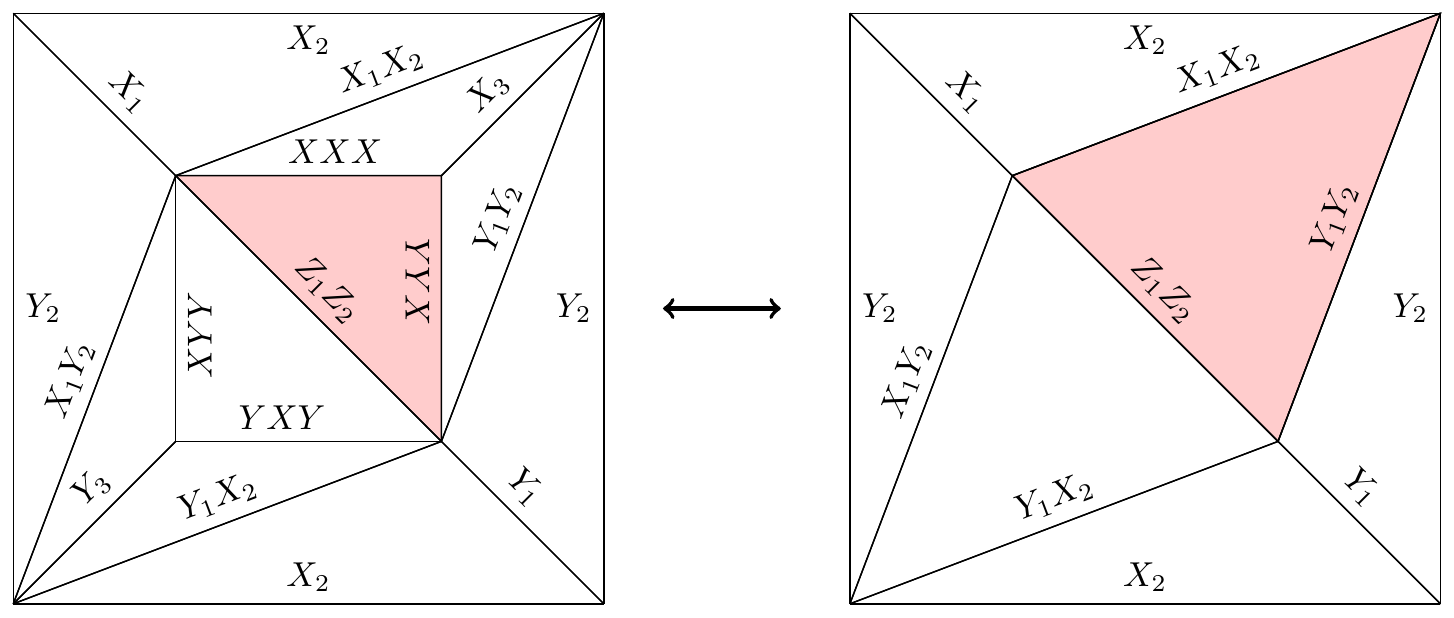}
\end{tabular}
\caption{\label{Pachner} Equivalence between Mermin's square and star. (a) Volume $V \in C_3(3)$ of Eq.~(\ref{SurfEq}). (b) Flipping between the surfaces $F_\text{star}$ (left) and $F_\text{square}$ (right), by adding the boundary $\partial V$. $\beta$ evaluates to one on the shaded faces. }
\end{center}
\end{figure}

The volume $V$ of Eq.~(\ref{SurfEq}) is depicted in Fig.~\ref{Pachner}a. The surfaces $F_\text{star}$ and $F_\text{square}$ are shown in Fig.~\ref{Pachner}b. They are obtained from another by adding the boundary $\partial V$. The Mermin square resulting from this procedure is locally rotated w.r.t. the standard convention, namely
$$
\includegraphics[width=3cm]{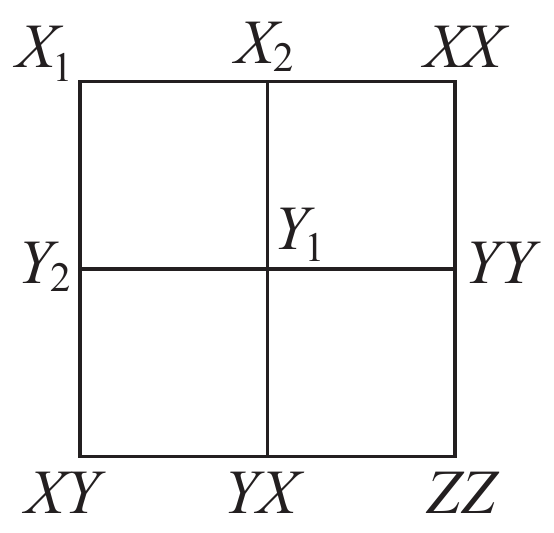}
$$

\subsection{State-dependent parity proofs}\label{sdpp}

Mermin's star---whose state-independent version was discussed in Section~\ref{CohPar}---also exists in a state-dependent version~\cite{Merm}. We use it as an initial  example, to illustrate the adaption of the topological argument to the state-dependent case and to motivate the definitions Eq.~(\ref{defS}) and Def.~\ref{DefSDvalue} below. The state-dependent Mermin star contains a special set 
$S=\{X_1X_2X_3, X_1Y_2Y_3, Y_1X_2Y_3, Y_1Y_2X_3\}
$ 
of observables and a special state, the Greenberger-Horne-Zeilinger state $|\text{GHZ}\rangle = (|000\rangle+|111\rangle)/\sqrt{2}$. The latter is a simultaneous eigenstate of the observables in $S$, with eigenvalues $+1,-1,-1,-1$, respectively. There is thus a value assignment $s(XXX)=0$, $s(XYY)=s(YXY)=s(YYX)=1$. From the perspective of non-contextual hidden variable models, the question is whether the value assignment $s$ can be extended in a consistent fashion to the local observables $X_i$ and $Y_i$.

\begin{figure}
\begin{center}
\begin{tabular}{lclcl}
(a) && (b) && (c)\\
\includegraphics[height=4.5cm]{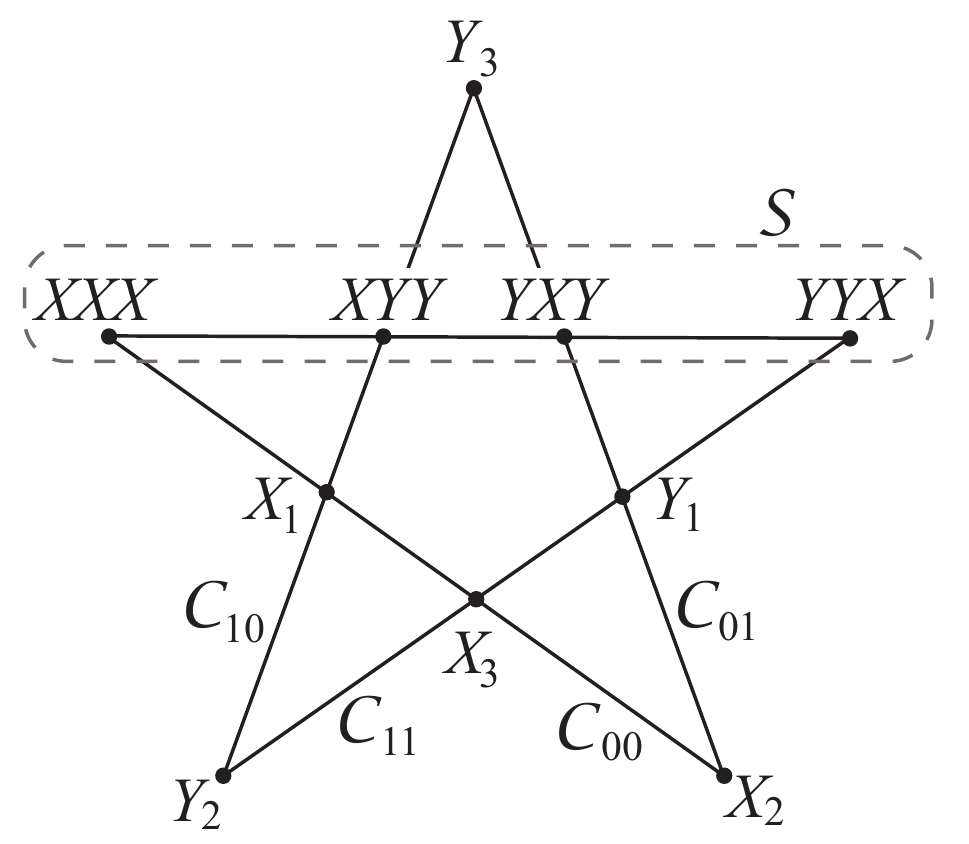} && \includegraphics[height=4.5cm]{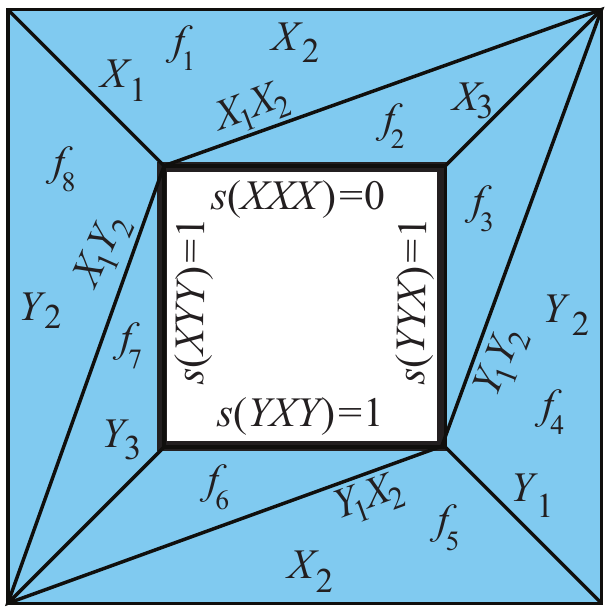} && \includegraphics[height=4.5cm]{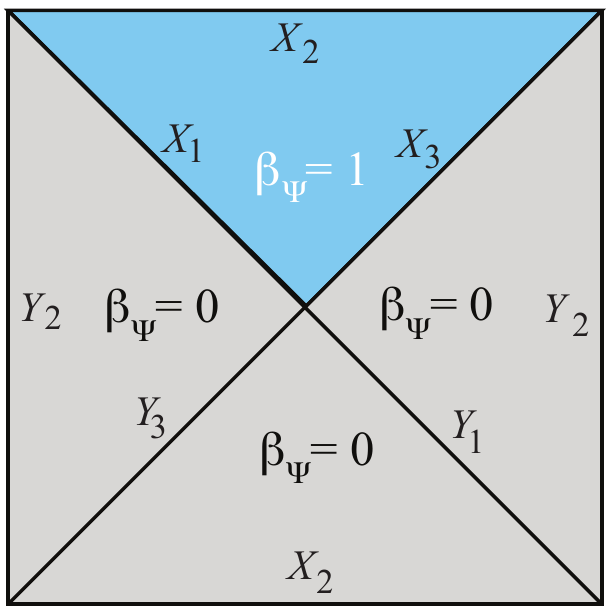}
\end{tabular}
\caption{\label{MSSD2}State-dependent version of Mermin's star. (a) One of the five contexts now defines a quantum state, in this case the Greenberger-Horne-Zeilinger state \cite{GHZ}. The other contexts, ${\cal{C}}_{00}$ .. ${\cal{C}}_{11}$, remain for measurement. (b) The state-dependent Mermin star embedded in a chain complex ${\cal{C}}$. (c) The state-dependent Mermin star embedded in the smaller complex ${\cal{C}}_*(E,E_\Psi)$ obtained from ${\cal{C}}_*(E)$ by contraction of the edges $E_\Psi$ corresponding to the GHZ-stabilizer.  The function $\beta_\Psi$ evaluates to $0$ on the face displayed in blue, and to $1$ on the three faces displayed in light gray.} 
\end{center}
\end{figure}

Adapting the topological state-independent argument, we now demonstrate that this is not the case. We choose the mapping $\eta$ such that $X_1X_2X_3,X_1Y_2Y_3, Y_1X_2Y_3,Y_1Y_2X_3,X_i,Y_i \in \eta(E)$, and consider the surface $F=\sum_{i=1}^8f_i$ displayed in Fig.~\ref{MSSD2}b. For any consistent value assignment $s$ we thus have $s(\partial F)=s(XXX)+s(XYY)+s(YXY)+s(YYX) \mod 2 =1$. On the other hand, $\beta(f_i)=0$, for $i=1,..,8$. Thus, assuming the existence of a consistent value assignment $s$, with $ds = \beta$ (cf. Lemma~\ref{ConsistBeta}) and with Stokes' theorem, we arrive at the following contradiction (addition $\text{mod}\; 2$):
$$
0 = \int_F \beta = \int_F ds = \int_{\partial F} s = 1.
$$ 
Hence our assumption that a consistent value assignment exists must be wrong.\medskip
 
We now turn to the general state-dependent scenario. 
Any state-dependent contextuality proof singles out a subset ${\cal{O}}_\Psi \subset {\cal O}$ of observables of which a special state $|\Psi\rangle$ is an eigenstate. Namely,
\begin{equation}\label{defS}
{\cal{O}}_\Psi := \left\{O \in {\cal{O}}|\; \exists \,s_O \in \mathbb{Z}_d \, \text{such that}\;O|\Psi\rangle = \omega^{s_O}|\Psi\rangle \right\}.
\end{equation}
The set ${\cal{O}}_\Psi$ may or may not be a context. It is required of ${\cal{O}}_\Psi$ that the observables therein have at least one joint eigenstate, $|\Psi\rangle$, but it is not required of them that they commute.

We want to integrate this extra bit of information into our topological description.  By the definition of ${\cal{O}}$ and Eq.~(\ref{defS}), the set ${\cal{O}}_\Psi$ has the property that whenever $[O_1,O_2]=0$ for $O_1,O_2\in {\cal O}_\Psi$ the product $O_1O_2$ also lies in ${\cal O}_{\Psi}$. We need this condition to be able to construct a subcomplex of ${\cal C}_*={\cal C}_*(E)$. 
The corresponding labels determine a subset $E_\Psi\subset E$ of edges and a subcomplex ${\cal C}_*(E_\Psi)$ whose definition is analogous to ${\cal C}_*$.  

Let us define $s_\Psi:E_\Psi \rightarrow \mathbb{Z}_d$ via Eq.~(\ref{defS}), i.e.,
\begin{equation}\label{sPsi}
T_a|\Psi\rangle = \omega^{s_\Psi(a)}|\Psi\rangle,\;\;\forall a \in E_\Psi.
\end{equation}
We can regard $s_\Psi$ as an element of $C^1(E_\Psi)$ by extending it linearly. A consistent value assignment in the state-dependent case has to be compatible with the eigenvalues   on the given state. This suggests the following definition.   
\begin{Def}\label{DefSDvalue}
A state-dependent consistent value assignment is a function $s:E\rightarrow \mathbb{Z}_d$ that satisfies
\begin{equation}\label{SDeq}
s(a)+s(b)-s(a+b)=\beta(a,b)
\end{equation}
for all commuting $(a,b)\notin E_\Psi\times E_\Psi$, and its restriction to $E_\Psi$ coincides with $s_\Psi$.
\end{Def}
According to Eq.~(\ref{SDeq}) only the commuting labels which are not contained in $E_\Psi$ matters. Geometrically we can remove the edges in $E_\Psi$ by contracting them. For the example of Mermin's star, this process is depicted in Fig.~\ref{MSSD2}. On the algebraic side, the chain complex of the contracted space is described by the relative complex  defined by the quotient 
$$
{\cal C}_*(E,E_\Psi) = {\cal C}_*(E)/{\cal C}_*(E_\Psi).
$$
In this quotient edges, the faces, and volumes which come from $E_\Psi$ are removed. Therefore we can think of this complex as having edges in the complement $E-E_\Psi$ of the set $E_\Psi$. More explicitly, a $1$-chain in this complex can be identified as a sum
$$
\sum_{a\in E-E_\Psi} \alpha_a[a] \;\; \text{ where } \alpha_a\in \mathbb{Z}_d
$$ 
similarly $2$-chains are linear combinations of  commuting elements not contained in $E_\Psi\times E_\Psi$. 
We refer to the boundary operator of ${\cal C}_*(E,E_\Psi)$ as the relative boundary operator and denote it by $\partial_R$ to distinguish it from $\partial$.
The boundary operator  $\partial_R$ is the same as $\partial$ except that the edges, faces or volumes corresponding to $E_\Psi$ are removed. For example, in Mermin's star of Fig.~\ref{MSSD2}b, $\partial_R (f_1+f_2)= a_{X_1}+a_{X_2}+a_{X_3}$, whereas $\partial (f_1+f_2)= a_{X_1}+a_{X_2}+a_{X_3}+a_{XXX}$. In general the relative boundary $\partial_R f$ of a $2$-chain $f$ is the sum of the edges in $\partial f$ which lie in $E-E_\Psi$.

The relation between the chain complexes we defined so far can be expressed as a short exact sequence  
\begin{equation*} 
0\rightarrow \mathcal{C}_*(E_\Psi) \rightarrow \mathcal{C}_*(E)\rightarrow \mathcal{C}_*(E,E_\Psi)\rightarrow 0
\end{equation*}
and the corresponding short exact sequence of cochain complexes is
\begin{equation*}
0\rightarrow \mathcal{C}^*(E,E_\Psi)\rightarrow \mathcal{C}^*(E)\rightarrow  \mathcal{C}^*(E_\Psi)\rightarrow 0.
\end{equation*}
Note that $\mathcal{C}^*(E,E_\Psi)$ can be characterized as  cochains in $\mathcal{C}^*$ whose restriction to $E_\Psi$ vanishes.
We will interpret Def.~\ref{DefSDvalue} using the cochain complex $\mathcal{C}^*(E,E_\Psi)$. In order to do  this $\beta$ must be modified so that it vanishes on all faces whose boundary is in $E_\Psi$. 
We will denote the modified function by $\beta_\Psi$, and show that it is a cocycle in  $C^2(E,E_\Psi)$. We define
\begin{equation}\label{Def_BetaPsi}
\beta_\Psi = \beta + d s_\Psi
\end{equation}
where $s_\Psi$ is regarded as a function $E\rightarrow \mathbb{Z}_d$ by defining it to be zero on  $E-E_\Psi$.

\begin{Theorem} \label{T1SD}
If $[\beta_\Psi]\neq 0$ in $H^2(\mathcal{C}(E,E_\Psi),\mathbb{Z}_d)$ then the pair $(\mathcal{O},|\Psi\rangle)$ exhibits state dependent contextuality.
\end{Theorem}

{\em{Proof of Theorem~\ref{T1SD}.}} 
Given a $2$-chain $f\in C_2(E)$ with boundary
$$
\partial f =\sum_{a\in E} \alpha_a[a]
$$ 
our definition yields
$$
\beta_\Psi(f) = \beta(f) + \sum_{a\in E_\Psi} \alpha_a s_\Psi(a).
$$
Note that $\beta_\Psi$ vanishes on faces whose boundary is in $E_\Psi$. To see this let $a,b\in E_\Psi$ be two commuting elements. 
Then,
$$
\begin{array}{rcl}
\beta_\Psi(a,b) &=& \beta(a,b) + (s_\Psi(a)+s_\Psi(b)-s_\Psi(a+b))\\
&=& - (s(a) +s(b)-s(a+b)) + (s_\Psi(a)+s_\Psi(b)-s_\Psi(a+b))\\
&=& 0.
\end{array}
$$
Therein, the first line is the definition of $\beta_\Psi$, Eq.~(\ref{Def_BetaPsi}). The second line follows by Lemma~\ref{ConsistBeta}, and the third line by the second item of Def.~\ref{DefSDvalue}.
As a result, $\beta_\Psi$ is an element of $C^2(E,E_\Psi)$. Moreover it is a cocyle since $d\beta_\Psi=d\beta+dds_\Psi=0$. The remainder of the proof proceeds as in  Theorem~ \ref{T1}. There is a $1$-cochain $s$ which satisfies  Def.~\ref{DefSDvalue} if and only if the cohomology class $[\beta_\Psi]$ vanishes.
$\Box$ \medskip

Finally, we return to our initial example of the state-dependent Mermin star, and explain it in terms of the relative cocycle $\beta_\Psi \in C^2(E,E_\Psi)$. Although the  new argument is almost exactly the same as the former (which used $\beta$ and $s_\Psi$), we give it here in order to invoke in an example the above-introduced notions of $\beta_\Psi$ and ${\cal{C}}_*(E,E_\Psi)$. The chain complex ${\cal{C}}_*(E,E_\Psi)$ corresponding to the state-dependent Mermin star has four elementary faces shown in Fig.~\ref{MSSD2}c. $\beta_\Psi$ evaluates to 1 on one of those faces, and to 0 on the other three. Thus, for the surface $F$ consisting of these four elementary faces, $\beta_\Psi(F) = 1$. We further have $\partial_{R} F=0$. 

Now assume that a consistent non-contextual value assignment $s$ exists, $\beta_\Psi =-ds$. Then, $1=\beta_\Psi(F)=ds(F)=s(\partial_{R}F)=s(0)=0$. Contradiction.

\section{Symmetry-based proofs of contextuality}\label{SbP}

The contextuality proofs in this section are based on invariance transformations. They lead the assumption of the existence of non-contextual value assignments into an algebraic contradiction, as did the parity-based proof encountered before. The new ingredient of these proofs is symmetry, and its representation in terms of group cohomology. 

The main results of this section are Theorems~\ref{C1b}, \ref{C1c}, \ref{C1bSD} and \ref{C1cSD} relating contextuality to the cohomology of the symmetry group. Also, we establish a relation between symmetry-based contextuality proofs and the parity-based proofs of Section~\ref{CohPar}; see Corollary~\ref{rel}.

\subsection{First example based on Mermin's square}\label{SBMS}

\begin{figure}
\begin{center}
\begin{tabular}{lclcl}
(a) && (b) && (c)\\
\includegraphics[height=3cm]{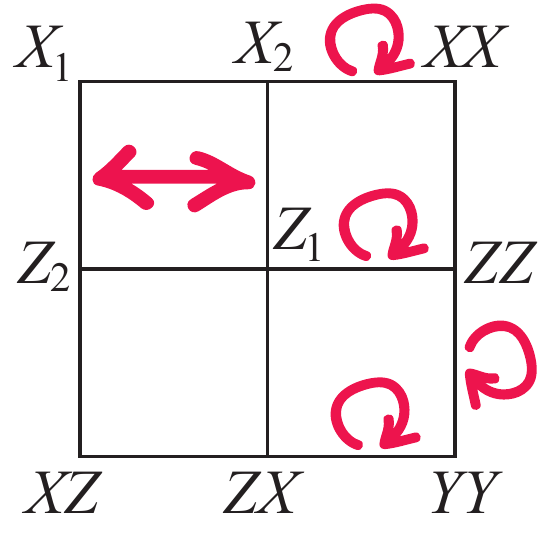} && \includegraphics[height=3cm]{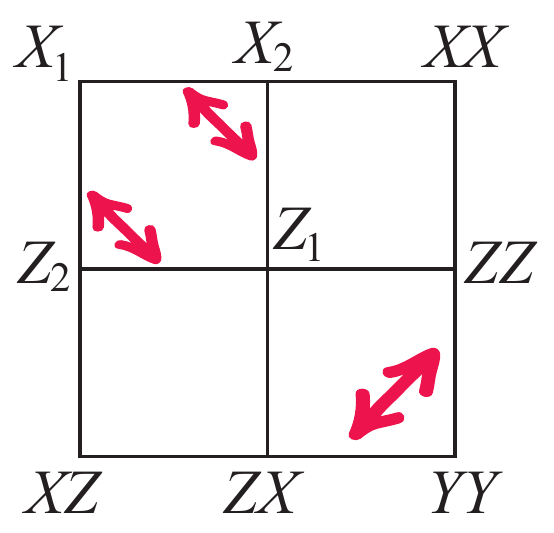} && \includegraphics[height=3cm]{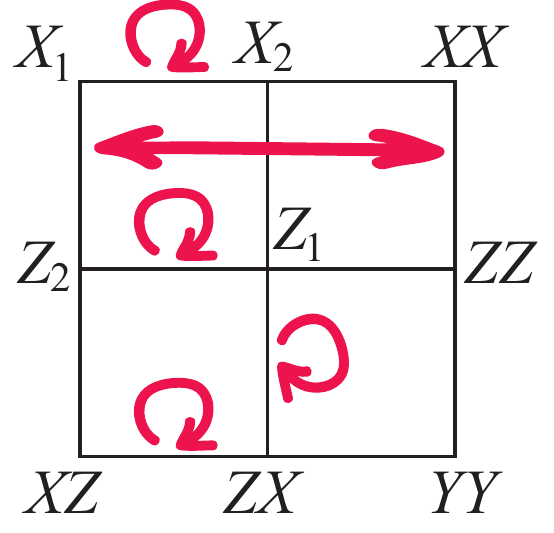} 
\end{tabular}
\caption{\label{SymmTrans}Symmetry transformations of Mermin's square. (a) Exchange of qubits 1 and 2, (b) Hadamard gate on qubit 1, (c) CNOT gate between qubits 1 and 2.}
\end{center}
\end{figure}

To illustrate the concept of contextuality proofs based on a symmetry $G$ of a set ${\cal{O}}$ of observables, we return to our earlier example of Mermin's square.
We find that it is invariant under certain symmetry transformations, for example the exchange of the two qubits, a Hadamard gate on qubit 1 or 2, or the CNOT gate between qubits 1 and 2; See Fig.~\ref{SymmTrans}. The square is mapped to itself under these transformations, with the observables in ${\cal{O}}$ and the contexts being permuted, and observables possibly flipping signs. Consider, in particular, the transformation of the square under the Hadamard gate $H_1$. In this case, the Pauli observable $Y_1Y_2$ changes its sign under conjugation, whereas all the other observables in the square map to one another without incurring sign changes. As we discuss now, a contextuality proof can be extracted from this transformation behaviour. This proof is of a different kind than the earlier parity proof, since the parity proof does not invoke any symmetry transformation.

For this example, $\eta(E)$ is 
$$
\eta(E) =\{I, X_1, X_2, X_1X_2,Z_1,Z_2,Z_1Z_2,X_1Z_2,Z_1X_2,Y_1Y_2 \},
$$
and $E$ is the corresponding index set. The Hadamard gate $H_1$ on the first qubit is in the symmetry group $G$ for Mermin's square, i.e. it maps the set $\mathcal{O} = \pm \eta(E)$ to itself. For example, $H_1: X_1 \leftrightarrow Z_1$, $Y_1Y_2\leftrightarrow -Y_1Y_2$, etc. The latter minus sign is important for the proof.

Assume that a consistent non-contextual value assignment $s$ exists. Then, from it, an new value assignment $s'$ can be constructed that is obtained from $s$ by application of the Hadamard gate $H_1$. Namely,
\begin{equation}\label{s_to_spr}
s'(a_{X_1}) =s(a_{Z_1}),\;s'(a_{Z_1}) =s(a_{X_1}),\;s'(a_{Y_1Y_2}) =s(a_{Y_1Y2}) + 1\mod 2,\;\; \text{etc}.
\end{equation}
We now consider the quantity
\begin{equation}\label{EtaDef}
\chi(s) = \sum_{a \in E} s(a) \mod 2.
\end{equation}
With the above transformation $s\longrightarrow s'$, we observe that
\begin{equation}\label{chi1}
\chi(s') =\chi(s) + 1 \mod 2.
\end{equation}
Now consider obtaining the value assignment $s'$ from $s$ by flipping individual values. To preserve the product constraints in the square---which from the perspective of contextuality are the relevant information contained in ${\cal{O}}$---there must be an even number of flips in every row and column of the square, and hence
\begin{equation}\label{chi2}
\chi(s) = \chi(s').
\end{equation} 
This is in contradiction to Eq.~(\ref{chi1}), and our assumption that a consistent value assignment $s$ existed must be wrong.

\subsection{The symmetries of ${\cal{O}}$}\label{SymmO}

For our general setting, we consider transformations $g\in G$ that satisfy the following two properties.
\begin{itemize}
\item[(i)]{The set ${\cal{O}}$ is preserved under all transformations in $G$. That is, there is an action of $G$ on ${\cal{O}}$ and an induced action of $G$ on $E$ such that
\begin{equation}\label{TTrans}
g(T_a) = \omega^{\tilde{\Phi}_g(a)}T_{ga}, \;\; \forall g\in G.
\end{equation}
Therein, $\tilde{\Phi}$ is the so-called phase function. It describes how observables in ${\cal{O}}$ transform under the symmetry group $G$.}
\item[(ii)]{Multiplication in all abelian subgroups of ${\cal{O}}$ is preserved,
\begin{equation}\label{Mpres}
g(O_1O_2) = g(O_1)g(O_2),
\end{equation}
for all pairs of commuting $O_1,O_2 \in {\cal{O}}$ and all $g\in G$.}
\end{itemize} 
The conjugation by a Hadamard gate $H_1$ on qubit 1 described in Section~\ref{SBMS}, $H_1(T_a)=H_1 T_a H_1^\dagger$, is a special case of the transformations Eq.~(\ref{TTrans}), (\ref{Mpres}).

The above transformations $g$ form a group under composition. Let $\Aut({\cal O})$ denote the group of all  symmetries of ${\cal O}$, that is all the transformations satisfying Eq.~(\ref{TTrans})-(\ref{Mpres}).
An action of a group $G$ as defined above gives a group homomorphism
\begin{equation}\label{rep}
G\rightarrow \Aut({\cal O})
\end{equation}
which sends a group element $g$ to the transformation determined by Eq.~(\ref{TTrans}).

Eq.~(\ref{TTrans}) can be understood as a coordinate transformation. Commuting observables obey the same algebraic relations before and after the transformation. The constraint Eq.~(\ref{Mpres}) enforces this property.

It is useful to restate Eq.~(\ref{Mpres}) in terms of $\eta(E) \subset {\cal{O}}$. It then reads
\begin{equation}\label{TTrans2}
g(T_{a+b}) = \omega^{\beta(a,b)}g(T_a)g(T_b),\;\; \forall g\in G, \;\; \text{for all commuting}\;  T_a,T_b \in {\cal{O}} .
\end{equation}
Thus, for all $g \in G$, the function $\beta: C_2\longrightarrow \mathbb{Z}_d$ is the same before and after the transformation.

The phase function $\tilde{\Phi}$ satisfies a further constraint resulting from the compatibility with the group structure of $G$. Namely, we require that $(gh)(T_a)=g(h(T_a))$, for all $g,h\in G$ and all $a\in E$. \medskip

To state the above two conditions in a convenient form, we develop further the underlying topological notions.
The function $\tilde{\Phi}$ assigns to a group element $g\in G$ a function $\tilde{\Phi}_g:C_1\rightarrow \mathbb{Z}_d$. Therefore we can think of $\tilde{\Phi}$ as an element of $C^1(G,C^1)$, the group of $1$-cochains which takes values in $C^1$.
We can also regard $\beta$ as an element of $C^0(G,C^2)$ by identifying $0$-cochains with the coefficient group $C^2$. To express the properties of $\tilde{\Phi}$ in a compact way we introduce the more general object $C^p(G,C^q)$. These are $p$-cochains on $G$ taking values in the group $C^q$ of  $q$-cochains in the complex $\cal{C}$. There are two types of differentials
\begin{equation}\label{VerHorDif}
\begin{CD}
C^p(G,C^{q+1})  @. \\
@Ad^vAA       @.     \\
C^p(G,C^q) @>d^h>>  C^{p+1}(G,C^q).
\end{CD}
\end{equation}
The vertical differential $d^v$ is induced by the differentials in $\cal{C}$, the horizontal differential $d^h$ is the group cohomology differential.  

\begin{Lemma}\label{PhiProp}
For all phase functions $\tilde{\Phi}$ defined through Eq.~(\ref{TTrans}) it holds that
\begin{subequations}\label{diff}
\begin{align}\label{diff_h}
d^h \tilde{\Phi} &=0,\\ 
\label{diff_v}
d^v\tilde{\Phi} &=d^h\beta.
\end{align}
\end{subequations}
\end{Lemma}
The cocycle $\beta$ and the phase function $\tilde{\Phi}$, along with its ``essence'' $\Phi$ introduced below, are the central physical objects in this paper. $\beta$ describes algebraic relations among commuting observables in ${\cal{O}}$, and $\tilde{\Phi}$ describes the transformation behaviour of these observables under the symmetry group $G$. Eq.~(\ref{diff}) shows that these two quantities are linked.\smallskip

{{\em{Proof of Lemma~\ref{PhiProp}.}} Regarding Eq.~(\ref{diff_h}),} with the transformation rule Eq.~(\ref{TTrans}) for observables, we find
$$
(gh) (T_a)  = \omega^{\tilde{\Phi}_{gh}(a)}T_{gh\,a},\;\; \forall a \in E,\,\forall g,h\in G.
$$
Alternatively, using group compatibility $(gh)(T_a)=g(h(T_a))$, we find $\forall a \in E,\,\forall g,h\in G$
$$
\begin{array}{rcl}
(gh)(T_a)  &=& g(h(T_a))\\
 &=& \omega^{\tilde{\Phi}_h(a)} g(T_{ha}) \\
 &=& \omega^{\tilde{\Phi}_h(a)} \omega^{\tilde{\Phi}_g(ha)} T_{gh\,a}.
\end{array}
$$
Comparing the two expressions, we find the group compatibility condition
\begin{equation}\label{Gcompat}
\tilde{\Phi}_h(a)+\tilde{\Phi}_g(ha) - \tilde{\Phi}_{gh}(a) =0, \;\; \forall g\in G,\, \forall a\in C_1,
\end{equation}
which is Eq.~(\ref{diff_h}).

Eq.~(\ref{diff_v}) is a consequence of Eq.~(\ref{TTrans2}). We have
$$
\omega^{\tilde{\Phi}_g(a+b)}T_{g(a+b)} = g(T_{a+b}) = \omega^{\beta(a,b)}g(T_{a})g(T_b) =\omega^{\beta(a,b)+\tilde{\Phi}_g(a)+\tilde{\Phi}_g(b)-\beta(ga,gb)}T_{g(a+b)},
$$
and after rearranging it we obtain Eq.~(\ref{diff_v}). $\Box$\medskip 

The symmetry-based contextuality proofs discussed in this section will employ the phase function $\tilde{\Phi}$. Lemma~\ref{ConsTrans} below is a first link between the phase function and consistent value assignments.
\begin{Lemma}\label{ConsTrans}
If $\textbf{s}:E \longrightarrow \mathbb{Z}_d$ satisfies the consistency constraints Eq.~(\ref{ValTest}) of Lemma~\ref{ConsistBeta}, then so does $\textbf{s}': E\longrightarrow \mathbb{Z}_d$ defined for any given $g \in G$ by 
\begin{equation}\label{spr}
s'(a):=s(ga)+\tilde{\Phi}_g(a), \;\; \forall a \in E.
\end{equation}
\end{Lemma}
This Lemma provides the formal justification for Eq.~(\ref{s_to_spr}) in the contextuality proof of Section~\ref{SBMS}, namely the transformation of the value assignment $s$ into a new assignment $s'$ under the Hadamard gate $H_1$. Recall that all addition involving the phase function is $\text{mod}\;d$.\medskip

{\em{Proof of Lemma~\ref{ConsTrans}.}} With Eqs. (\ref{spr}), (\ref{diff_v}) and (\ref{ValTest}) we have
$$
\begin{array}{rcl}
s'(a)+s'(b)-s'(a+b) &=& s(ga) + s(gb) -s(g(a+b)) + \tilde{\Phi}_g(a) +  \tilde{\Phi}_g(b) -  \tilde{\Phi}_g(a+b)\\
&=& s(ga) + s(gb) -s(g(a+b)) + \beta(ga,gb) -\beta(a,b)\\
&=& -\beta(a,b).
\end{array}
$$
Thus, the same constraints Eq.~(\ref{ValTest}) satisfied by $s$ are also satisfied by $s'$. $\Box$ \medskip

\subsection{The general state-independent case}\label{GC}

Here we generalize the symmetry-based proof for Mermin's square given in Section~\ref{SBMS} to general sets ${\cal{O}}$ of observables with a sufficiently large symmetry group $G$. To begin, let's analyze the inner workings of that proof.

First, consider the sum $\chi(s)$ of value assignments. In cochain notation it reads $\chi(s)=s(e)$, for some 1-chain $e$ (in the above case, $e=\sum_{a\in E}\alpha_a[a]$ where $\alpha_a\in \mathbb{Z}_d$). In order to permit the comparison of Eq.~(\ref{chi1}), i.e., in order to have the same summation on the lhs and rhs, the transformation $g$ ($g=H_1$ in the proof of Section~\ref{SBMS}), needs to satisfy
\begin{equation}\label{eTrans}
g e = e.
\end{equation}
Further, in order to have definite values on either side of Eq.~(\ref{chi2}), $\chi(s)=s(e)$ needs to be a sum of constraints. In topological notation, we thus require that
\begin{equation}\label{fbdy}
e=\partial f,
\end{equation}
for some $f \in C_2$.

Finally, in order to have disagreement between the comparisons of Eq.~(\ref{chi1}) and (\ref{chi2}), we must require that
\begin{equation}\label{nontriv}
\tilde{\Phi}_g(e) \neq 0.
\end{equation}
The conditions Eq.~(\ref{eTrans}) - (\ref{nontriv}) are the central ingredients for the symmetry-based proofs. This leads us to the following result.

\begin{Lemma}\label{L4}
Given a set ${\cal{O}}$ of observables and the corresponding symmetry group $G$, if there exist a $g \in G$ and an $f \in C_2$ such that $g\,\partial f = \partial f$ and $\tilde{\Phi}_g(\partial f)\neq 0$ then ${\cal{O}}$ has state-independent contextuality.
\end{Lemma}

In addition to the above argument, we now give a formal proof for this Lemma.\\
{{\em{Proof of Lemma~\ref{L4}.}} Eq.~(\ref{diff_v}) implies that
$$
\tilde{\Phi}_g(\partial f) = d^v\tilde{\Phi}(g,f)=d^h\beta(g,f).
$$
Now under the assumption that there exists a value assignment $s$ satisfying Eq.~(\ref{ValTest}) and there exists $g \in G$ and $f \in C_2$ such that $g\partial f=\partial f$ this equation becomes
$$
\tilde{\Phi}_g(\partial f)=d^h\beta(g,f)=-d^hd^vs(g,f)=-(s(g\partial f) -s(\partial f)) =0.
$$
Therefore if $\tilde{\Phi}_g(\partial f) \not= 0$ we get a contradiction. $\Box$
}\medskip

\begin{figure}
\begin{center}
\begin{tabular}{lcl}
(a) && (b)\\
\includegraphics[height=4.4cm]{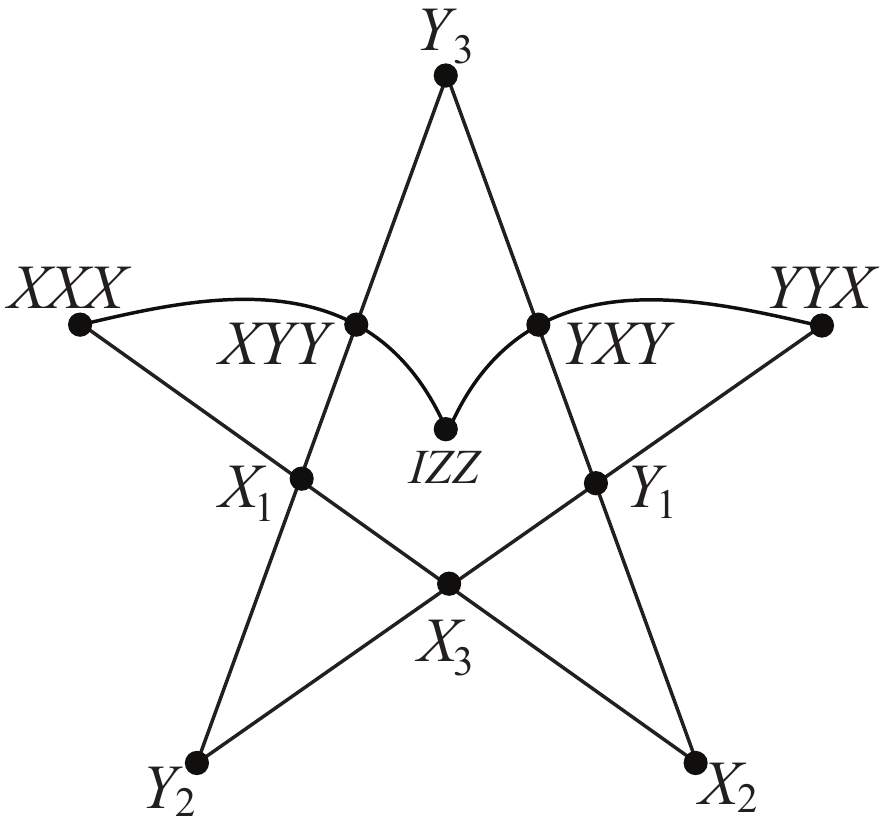} &&  \includegraphics[height=4.4cm]{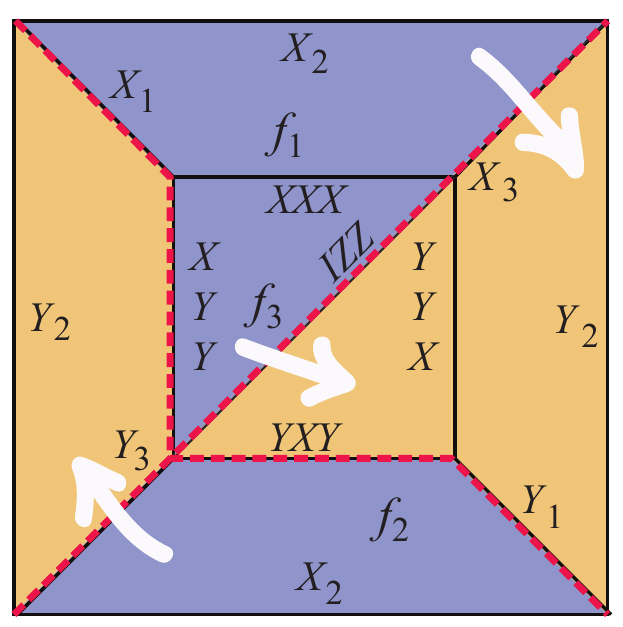}
\end{tabular}
\caption{\label{2MS} (a) Decorated Mermin star. (b) Topological representation of the decorated star. The left and right edges and the top and bottom edges, respectively, are identified. The three blue faces form $f$, $f=f_1+f_2+f_3$, and the orange faces form $A_1A_2 f$. Each white arrow points from a face $f_i$ to the corresponding face $A_1A_2 \,f_i$. The boundaries of $f$ and $A_1A_2 f$ coincide (dashed lines).}
\end{center}
\end{figure}

{\em{Example: decorated Mermin star.}} We present a symmetry-based proof for the ``decorated Mermin star'', depicted in Fig.~\ref{2MS}a, based on the symmetry transformation
\begin{equation}\label{SZ2sq}
g =  A_1 \otimes A_2 \otimes I_3,
\end{equation}
where $A:=(X+Y)/\sqrt{2}$. 
We call this version of Mermin's star ``decorated'', because of the additional observable $I_1Z_2Z_3$ which is not included in the original star, but automatically included in the corresponding setting derived from a complex ${\cal{C}}$ (cf. the first property of ${\cal{C}}$ described in Section~\ref{Complex}). This additional observable is of importance for the symmetry-based proof.
 
We show that for $g=A_1A_2$ the two conditions of Lemma~\ref{L4}, namely $\exists f \in C_2$ such that $A_1A_2\, \partial f = \partial f$ and $\tilde{\Phi}_{A_1A_2}(\partial f) \neq 0$ are met. Choose $f=f_1+f_2+f_3$, with 
$$
\begin{array}{rcl}
\partial f_1 &=& a_{X_1}+ a_{X_2}+a_{X_3}+a_{XXX},\\ 
\partial f_2 &=& a_{Y_1}+ a_{X_2}+a_{Y_3}+a_{YXY},\\ 
\partial f_3 &=& a_{XXX}+ a_{XYY} + a_{IZZ}.
\end{array}
$$
See Fig.~\ref{2MS}b for illustration. It is now easily verified that $\partial f = A_1A_2\,\partial f$. Furthermore, since $AZ=-ZA$, it holds that $\tilde{\Phi}_{A_1A_2}(a_{IZZ})=1$. For all other edges $a$ displayed in Fig.~\ref{2MS}b, it holds that $\tilde{\Phi}_{A_1A_2}(a)=0$. Finally, since $a_{IZZ} \in \{\partial f\}$, it follows that $\tilde{\Phi}_{A_1A_2}(\partial f)=1\neq 0$. The conditions of Lemma~\ref{L4} are thus met, and the decorated Mermin star is contextual. 

\subsection{Topological formulation}\label{TopForm}

We now reformulate Lemma~\ref{L4} in terms of cohomology groups, which are invariant objects in topology. The result is Theorem~\ref{C1b}. To this end, we investigate the effect of the transformations Eq.~(\ref{EtaTrans}) on $\tilde{\Phi}$. Changing from the map $\eta$ of Eq.~(\ref{etaDef}) to $\eta_\gamma$ induces the change
$$
\tilde{\Phi}_g(a) \longrightarrow \tilde{\Phi}'_g(a) = \tilde{\Phi}_g(a)+\gamma(a)-\gamma(ga),\;\; \forall g\in G,\, \forall a \in C_1.
$$
The cohomological interpretation of this equation is
\begin{equation}
\tilde{\Phi} \longrightarrow \tilde{\Phi}- d^h\gamma.
\end{equation}
The change of the map $\eta$ has no effect on contextuality, as was demonstrated in Section~\ref{beta}. The phase functions $\tilde{\Phi}$ thus group into equivalence classes
$$
[\tilde{\Phi}] = \{ \tilde{\Phi} - d^h\gamma, \forall \gamma\}.
$$
Together with Eq.~(\ref{diff_h}), this implies that $[\tilde{\Phi}] \in H^1(G,C^1)$.

To make contact with Lemma~\ref{L4}, we now restrict the 1-chains of $C_1$ on which the phase functions $\tilde{\Phi}_g$ are evaluated. The boundaries $B_1$ are contained in $C_1$ as a subgroup. We can write this as a short exact sequence
$$
0\rightarrow B_1\rightarrow C_1\rightarrow C_1/B_1\rightarrow 0.
$$
Now taking the duals of each group in this sequence gives an other short exact sequence. That is applying $\text{Hom}(-,\mathbb{Z}_d)$ to each group in the above exact sequence gives 
\begin{equation}\label{exatchain}
0\rightarrow V \rightarrow C^1\rightarrow U\rightarrow 0
\end{equation}
where $V=\text{Hom}(C_1/B_1,\mathbb{Z}_d)$ and $U=\text{Hom}(B_1,\mathbb{Z}_d)$.
More explicitly, $U$ consists of $\mathbb{Z}_d$-linear maps $B_1\rightarrow \mathbb{Z}_d$ and  $V$ is the set of 1-cocycles, i.e., the set of  1-cochains that vanish on boundaries,
$$
V=\{\textbf{v} \in C^1|\; d^v\textbf{v} =0\}.
$$
Let $\tilde{\Phi}|_{B_1}:G\rightarrow U$ denote the composition of $\tilde\Phi:G\rightarrow C^1$ with the map $C^1\rightarrow U$ in the short exact sequence in (\ref{exatchain}).

We still have the constraint
$$
d^h \tilde{\Phi}|_{B_1} = 0,
$$
and  re-parametrizing by $\gamma$ has the effect of
$$
\tilde{\Phi}|_{B_1} \mapsto \tilde{\Phi}|_{B_1} - d^h\gamma|_{B_1},
$$
and therefore
\begin{equation}
[\tilde{\Phi}|_{B_1}] \in H^1(G,U).
\end{equation}
We then have the following topological reformulation of Lemma~\ref{L4}.
\begin{Lemma}\label{C1}
For a given set ${\cal{O}}$ of observables and corresponding symmetry group $G$, if $[\tilde{\Phi}|_{B_1}] \neq 0 \in H^1(G,U)$ then ${\cal{O}}$ exhibits state-independent contextuality.
\end{Lemma}

{\em{Proof of Lemma~\ref{C1}.}} {The elements of $B_1$ are of the form $\partial f$ for some $2$--chain $f$.
By Lemma \ref{PhiProp} we have
$$
\tilde{\Phi}_g(\partial f) = d^v\Phi(g,f)=d^h\beta.
$$
As shown in the proof of Lemma \ref{L4} if there is a value assignment $s$, that is $d^vs=-\beta$, then
$$
\tilde{\Phi}_g(\partial f) = d^h\beta = d^h(-d^vs(g,f))=-d^hs(g,\partial f)
$$
where $d^vs(g,f)=s(g,\partial f)$ by definition of the horizontal differential.
In other words $\tilde{\Phi}|_{B_1}$ is the coboundary of $s|_{B_1}$ with respect to the group cohomology differential. So existence of a value assignment implies that $[\tilde{\Phi}|_{B_1}]=0$. $\Box$
}\medskip

We proceed to establish a further reformulation of Lemma~\ref{L4}, Theorem~\ref{C1b} below. It makes explicit the structure of the symmetry group $G$, which is of relevance for MBQC.
Namely, the symmetry group $G$ has a subgroup $N$ which fixes the edges. That is $n(T_a)= \omega^{\tilde\Phi_n(a)} T_a$ for all $n\in N$. We now make two observations:

(i) $N$ is normal in $G$. Hence  the set of equivalence classes $\{g n,\; n\in N\}$ forms a group $Q:=G/N$.

(ii) A symmetry-based contextuality proof according to Lemma~\ref{L4} works for a group element $g \in G$ if and only if it works for any $g n$, with $n\in N$. That is, symmetry-based contextuality proofs are properties of equivalence classes $\{g n,\;n\in N\}$, or, equivalently, of elements $q \in Q$.\medskip

A proof of statement (ii) is as follows. We verify that the conditions of Lemma~\ref{L4} are met for the pair $(g,f)$ if and only if they are met for the pair $(gn,f)$, with $n \in N$. We observe that $n\, a = a$, for all $n \in N$ and all $a \in E$. Thus, first, $g \partial f = \partial f\Longleftrightarrow gn \,\partial f = \partial f$.

Furthermore, by Eq.~(\ref{diff_v}) and since $nf=f$, it holds that 
$
\tilde{\Phi}_n(\partial f)=d^v\tilde\Phi (n,f)=d^h\beta(n,f)=  \beta(nf)-\beta(f)=0
$. Then, by group compatibility Eq.~(\ref{Gcompat}), $\tilde{\Phi}_{gn}(\partial f) = \tilde{\Phi}_g(\partial f)$.
Thus, second, $\tilde{\Phi}_{gn}(\partial f) \neq 0 \Longleftrightarrow \tilde{\Phi}_g(\partial f)\neq 0$. $\Box$\medskip

Let $\pi:G\rightarrow Q$ denote the quotient map and $\theta:Q\rightarrow G$ be a section of $\pi$ i.e. $\pi\theta(q)=q$ for all $q\in Q$. We define $\Phi:Q\rightarrow U$ to be the composition of $\tilde\Phi|_{B_1}:G\rightarrow U$ with $\theta$.
Then the observation that $\tilde\Phi_{gn}(\partial f) = \tilde\Phi_{g}(\partial f)$ for all $n\in N$ can be written as 
$$
\tilde\Phi|_{B_1}(g,\partial f)=\Phi(q,\partial f)
$$
where $q=\pi(g)$. Moreover, this observation combined with Eq.~(\ref{diff_h}) in Lemma \ref{PhiProp} implies that $\Phi$ is a cocycle.
\begin{Theorem}\label{C1b}
For a given set ${\cal{O}}$ of observables and corresponding symmetry group $G$, if $[\Phi]\neq 0 \in H^1(Q,U)$ then ${\cal{O}}$ exhibits state-independent contextuality.
\end{Theorem} 
This is our final result on symmetry-based contextuality proofs for the state-independent case.\medskip

{\em{Proof of Theorem~\ref{C1b}.}} By Lemma \ref{C1} we need to show $[\Phi]\neq 0 $ if and only if $[\tilde\Phi|_{B_1}]\neq 0$. By definition of  $\Phi$, its class $[\Phi]$ maps to $[\tilde\Phi|_{B_1}]$ under the map
$$
\pi^*:H^1(Q,U)\rightarrow H^1(G,U) 
$$ 
induced by the homomorphism $\pi:G\rightarrow Q$. That is, $[\tilde\Phi|_{B_1}]\neq 0$ implies $[\Phi]\neq 0$. For the converse assume $\tilde\Phi|_{B_1}$ is a coboundary:
$$
\tilde\Phi|_{B_1}(g,\partial f)= d^hs(g,\partial f)
$$
for some  $s:B_1\rightarrow \mathbb{Z}_d$. Since $\theta(q)\partial f= g\,\partial f$ for $q=\pi(g)$ and by definition of $\Phi$ we have
$$
\Phi(q,\partial f)=\tilde\Phi|_{B_1}(g,\partial f)= d^hs(g,\partial f).
$$
Therefore $[\Phi]=0$ in $H^1(Q,U)$. In other words $[\Phi]\neq 0$ implies $[\tilde\Phi|_{B_1}]\neq 0$. $\Box$

\subsection{Relation between parity-based and symmetry-based proofs}\label{ProofRel}

We have so far found two topological methods to prove Kochen-Specker theorems in the state-independent case, one involving the second cohomology group $H^2({\cal{C}},\mathbb{Z}_d)$ in a chain complex ${\cal{C}}$ and the other involving the first cohomology group $H^1(G,U)$ of a symmetry group $G$. In this section we show that these proofs are related. It turns out that the symmetry-based proofs are at most as strong as the parity proofs. A proof of the former kind always implies a proof of the latter kind.

\begin{Cor}\label{rel}
Every symmetry-based proof of contextuality implies a parity-based proof.
\end{Cor}

{\em{Proof of Corollary~\ref{rel}.}}{
What we actually proved in Lemma~\ref{C1} is $[\beta]=0$ implies $[\tilde{\Phi}|_{B_1}]=0$. {The other way around,} $[\tilde{\Phi}|_{B_1}]\not=0$ implies $[\beta]\not=0$.  $\Box$
}

\subsection{Contextuality and the group extension problem}

The group extension problem is concerned with the following question: ``Given two groups $Q$ and $N$, with an action of $Q$ on $N$, what are the groups $G$  such that $N\subset G$ and $Q=G/N$?''. Any such group $G$ is called an extension of $N$ and $Q$, which is expressed as a short exact sequence 
\begin{equation}\label{gExt}
0\rightarrow N \rightarrow G\rightarrow Q \rightarrow 0.
\end{equation}
The simplest way to compose the groups $Q$ and $N$ is via the semi-direct product, $G = Q\ltimes N$, but often  there are additional possibilities. A semi-direct product is a twisted version of the direct product $Q\times N$   i.e. when multiplying two elements
$$
(q_1,n_1)(q_2,n_2)=(q_1q_2,a_{q_2}(n_1)n_2)
$$
on the second factor $n_1$ is changed by an automorphism   which depends on $q_2$. 

For example, the quaternion group $Q_8$ has a normal subgroup $\mathbb{Z}_4$ and a quotient $\mathbb{Z}_2$, but $Q_8 \neq \mathbb{Z}_2 \ltimes \mathbb{Z}_4$, which can be seen by counting the elements of order two.

The structure of the group extension has implications on the detection of contextuality by  cohomology groups, as we now explain. The exact sequence (\ref{exatchain}) gives a short exact sequence of cochain complexes
$$
 0\rightarrow C^*(Q,V)\stackrel{i}{\rightarrow} C^*(Q,C^1) \stackrel{j}{\rightarrow} C^*(Q,U)\rightarrow 0
$$
which gives long exact sequence of cohomology groups
\begin{equation}\label{longexact}
\cdots\rightarrow H^1(Q,V)\stackrel{i}{\rightarrow} H^1(Q,C^1)\stackrel{j}{\rightarrow} H^1(Q,U)\stackrel{\sigma}{\rightarrow} H^2(Q,V)\rightarrow H^2(Q,C^1)\rightarrow \cdots
\end{equation}
see \cite[Proposition 6.1]{Brown}. In general $\sigma([\alpha])$ is defined by  lifting the cocycle $\alpha$ in $C^k(Q,U)$ to an element of $C^k(Q,C^1)$ which we denote by $\alpha'$, and then applying the (group cohomology) differential $d^h$. Then the coboundary $d^h\alpha'$ is in $C^{k+1}(Q,C^1)$. Its image under $j$ is zero since $j(d^h\alpha')=d^h(j(\alpha'))=d^h\alpha=0$.
Therefore $d^h\alpha'$  actually lies in $C^{k+1}(Q,V)$. The map $\sigma$ sends $[\alpha]$ to $[d^h\alpha']$. 
Now let us describe the class $\sigma([\Phi])$. 
 Recall that $\Phi:Q\rightarrow U$ is defined by the composition 
$Q\stackrel{\theta}{\rightarrow}G\stackrel{\tilde\Phi|_{B_1}}{\rightarrow} U$. 
As the lift of this class we can take $\Phi':Q\rightarrow C^1$ defined by the composition $Q\stackrel{\theta}{\rightarrow}G\stackrel{\tilde\Phi}{\rightarrow }C^1$. Then $j(\Phi')=\Phi$. Therefore we have 
\begin{equation}\label{sigma}
\sigma([\Phi])=[d^h\Phi'].
\end{equation}

\begin{Theorem}\label{C1c}
For a given set ${\cal{O}}$ of observables and corresponding symmetry group $G$, if $\sigma([\Phi])\neq 0 \in H^2(Q,V)$ then ${\cal{O}}$ exhibits state-independent contextuality.
\end{Theorem} 

{\em{Proof of Theorem~\ref{C1c}.}} If $\sigma([\Phi])\neq 0$ then the class $[\Phi]$ which maps to it cannot be zero. Now Theorem \ref{C1b} implies that ${\cal{O}}$ exhibits state-independent contextuality. $\Box$ \medskip

The gist of the above Theorems~\ref{T1} -- \ref{C1c} is thus the chain of implications
$$
\sigma([\Phi]) \neq 0 \in H^2(Q,V) \Longrightarrow [\Phi] \neq 0 \in H^1(Q,U) \Longrightarrow [\beta]\neq 0 \in H^2({\cal{C}},\mathbb{Z}_d) \Longrightarrow \text{${\cal{O}}$ is contextual}.
$$
Thus, $\sigma([\Phi])$, $[\Phi]$, $[\beta]$ are successively stronger contextuality witnesses.\medskip

We conclude this section by showing that the  weakest of these witnesses,  $\sigma([\Phi])$, is indeed strictly weaker than $[\Phi]$. We demonstrate this by example. First, the following observation is helpful.
\begin{Lemma}\label{split}
If $G=Q\ltimes N$ then $\sigma([\Phi])=0$.
\end{Lemma}
{\em{Remark:}} Lemma~\ref{split} can be strengthened to an ``if and only if'' if the symmetry group $G$ is large enough. See Lemma~\ref{converse} in Appendix~\ref{C}. \medskip

{\em{Proof of Lemma~\ref{split}.}} The proof essentially follows from Eq.~(\ref{sigma}). We can choose the section $\theta:Q\rightarrow G$ to be a group homomorphism since $G$ splits as a semi-direct product. Then $\theta$ induces a map $\theta^*:C^2(G,C^1)\rightarrow C^2(Q,C^1)$ of chain complexes. In this case we have $\Phi'=\theta^*(\tilde\Phi)$. By Eq.~(\ref{sigma}) we have $\sigma([\Phi])=[d^h\Phi']=[d^h\theta^*(\tilde\Phi)]=[\theta^*(d^h\tilde\Phi)]=0$. $\Box$ \medskip
 
Thus, if $G=Q\ltimes N$ we do not have any hope for detecting contextuality by the cohomology class  $\sigma([\Phi]) \in H^2(Q,V)$. We use this observation in the example of the decorated Mermin star, discussed at the end of Section~\ref{GC}. Consider as the symmetry group $G$ the group generated by $u(g)=A_1A_2I_3$ (cf. Eq.~(\ref{SZ2sq})) and the set of all 3-qubit Pauli operators, ${\cal{P}}_3$. The Pauli operators form the normal subgroup $N$, and $Q=G/N \cong \mathbb{Z}_2$. Note that $g^2=I$, and 
$$
\mathbb{Z}_2 \ni 0 \mapsto I,  \mathbb{Z}_2 \ni 1 \mapsto g.
$$
We have in particular that $\langle g\rangle \cap ({\cal{P}}_3=N) =I$, and thus $G = \mathbb{Z}_2 \ltimes N$. This means $\sigma([\Phi])=0$ by Lemma \ref{split}, and the symmetry group under consideration does not provide a contextuality proof via Theorem~\ref{C1c}.
Yet, $[\Phi]\not=0$ since $[\tilde\Phi|_{B_1}]\not=0$, and a contextuality proof is  provided by Theorem~\ref{C1b}.\medskip

\subsection{State-dependent contextuality proofs based on symmetry}\label{SbPsd}

State-dependent, symmetry-based proofs of contextuality have previously been constructed by Spekkens, Edwards and Coecke \cite{SC} and by J. Lawrence \cite{JLaw}, for GHZ-scenarios. Here we describe general such contextuality proofs, and relate them to group cohomology. 

\begin{figure}
\begin{center}
 \includegraphics[height=6cm]{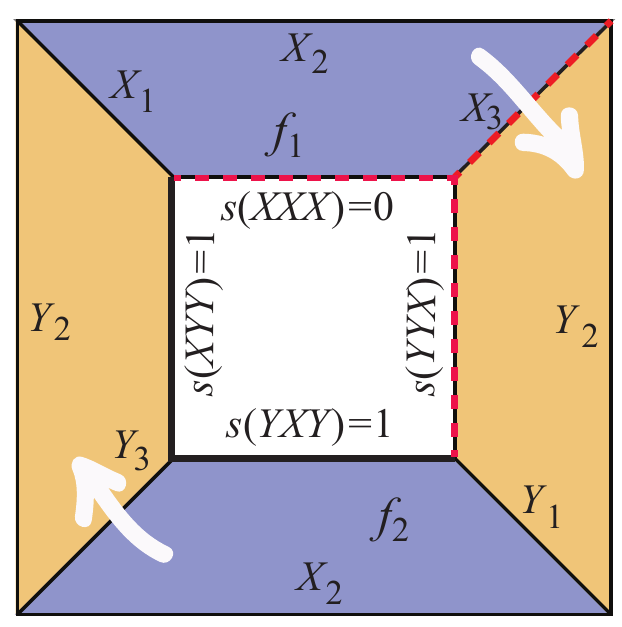}
\caption{\label{MSSD}Symmetry-based contextuality proof for the state-dependent version of Mermin's star, embedded in a chain complex ${\cal{C}}_*(E)$. The arrows map  faces $f_i$ to $A_1A_2Y_3\,f_i$. The observables $T_a$ corresponding to edges $a$ displayed dashed are flipped to $-T_a$ under conjugation by $A_1A_2Y_3$.} 
\end{center}
\end{figure}

The symmetry group $G$ of the state-independent case preserves $E$ and $\beta$.  It is now replaced by a subgroup $H \subset G$ which preserves $E$, $E_\Psi$, $\beta$ and $s_\Psi$. For any $g\in G$ that preserves $E_\Psi$ the action on $s_\Psi$ is as follows. The resource state $|\Psi\rangle$ is an eigenstate of any $T_a$, $a\in E_\Psi$, with eigenvalue $\omega^{s_\Psi(a)}$. Thus, $\langle T_a\rangle_\Psi = \omega^{s_\Psi(a)}$, for all $a \in E_\Psi$. By Eq.~(\ref{TTrans}), under the transformation $g$ the expectation value $\langle T_a\rangle_\Psi$ transforms as $\langle T_a\rangle_\Psi \longrightarrow \langle g(T_a)\rangle_\Psi = \omega^{\tilde{\Phi}_g(a)}\omega^{s_\Psi(ga)} = \omega^{s_\Psi'(a)}$. Hence, the update rule for the values $s_\Psi$ is
$$
s_\Psi(a) \longrightarrow s_\Psi'(a) = s(ga) + \tilde{\Phi}_g(a),
$$
for all $g\in G$ such that $g(E_\Psi)=E_\Psi$ and all $a \in E_\Psi$. Now the extra condition on the subgroup $H \subset G$ is that $s_\Psi'\equiv s_\Psi$, for all $h \in H$. Thus, in topological notation,
\begin{equation}
\tilde{\Phi}_h(a) = - d^hs_\Psi(h,a),\; \forall a\in E_\Psi,\,\forall h \in H. 
\end{equation}
It is useful to illustrate these symmetry constraints with the example of the state-dependent Mermin star; See Fig.~\ref{MSSD}. We consider the transformation $g \in G$ that has a unitary projective representation $u(g)=A_1A_2I_3$, which acts on the observables in ${\cal{O}}$ by conjugation. It preserves $E$ and $\beta$ as we have seen before, and it also preserves $E_\Psi$. But it does not preserve $s_\Psi$. For example, since $\tilde{\Phi}_g(a_{XXX})=0$, it holds that $s'_\Psi(a_{XXX}) =s_\Psi(a_{YYX}) + 0 =1$, whereas $s_\Psi(a_{XXX})=0$. Likewise, $s'_\Psi(a_{YYX})=0$ but $s_\Psi(a_{YYX})=1$. The values of $s_\Psi(a_{XXX})$ and $s_\Psi(a_{YYX})$ are thus flipped, while the values $s_\Psi(a_{XYY})$ and $s_\Psi(a_{YXY})$ remain unchanged.
Therefore, $g \not \in H$. 

However, we can find a related transformation $h \in H$, defined via $u(h)=Y_3 u(g) = A_1A_2Y_3$. Namely, the extra operation $Y_3 $ flips $s'_\Psi(a_{XXX})$ and $s'_\Psi(a_{YYX})$ back, and leaves $s'_\Psi(a_{YXY})$ and $s'_\Psi(a_{XYY})$ unaffected. The action of $Y_3$ also preserves $E$, $E_\Psi$ and $\beta$. In total, $h$ preserves $E$, $E_\Psi$, $\beta$ and $s_\Psi$, and is thus in the symmetry group $H$.

We will formulate symmetry based state-dependent contextuality proofs using the symmetry group $H$ and the relative complex ${\cal C}_*(E,E_\Psi)$.  The symmetry group $H$ preserves $E_\Psi$ by definition. It acts on the chain complex ${\cal C}_*(E_\Psi)$ by permuting the edges, faces, and volumes in each dimension. There is an induced action on the quotient ${\cal{C}}_*(E,E_\Psi)$. Geometrically we can think of this action as the permutation of the cells of the contracted space. The action on the chains gives an action on the cochains.  By replacing $G$ and ${\cal C}_*(E)$ in Diag.~(\ref{VerHorDif}) by $H$ and  ${\cal C}_*(E,E_\Psi)$ we consider the cochain  complex $C^p(H,C^q(E,E_\Psi))$ with horizontal $d^h$ and vertical $d^v$ differentials.
As before $d^h$ is induced by the group cohomology differential, and $d^v$ is induced by the relative boundary operator. The counterpart of Lemma~\ref{L4} for the state-dependent case is the following.

\begin{Lemma}\label{L7}
Given a set ${\cal{O}}$ of observables, a quantum state $|\Psi\rangle$ and the corresponding symmetry group $H$, if there exists an $h \in H$ and an $f \in C_2$ such that $h\,\partial_R f = \partial_R f$ and $\tilde{\Phi}_h(\partial_R f)\neq 0$ then ${\cal{O}}$ has state-dependent contextuality.
\end{Lemma} 

{\em{Example:}} The prototypical state-dependent contextuality scenario is the state-dependent version of Mermins's star \cite{Merm}, depicted in Fig.~\ref{MSSD}. In this case, the set $S=\{XXX,XYY,YXY,YYX\}$ is a context. The state $|\Psi\rangle$ is the Greenberger-Horne-Zeilinger (GHZ) state $|\text{GHZ}\rangle=(|000\rangle+|111\rangle)/\sqrt{2}$ \cite{GHZ}. The symmetry group $H$ is generated by permutation  of the three particles, the transformation $A_1\otimes A_2 \otimes Y_3$, and the GHZ stabilizer $S$. Choose $H\ni h=A_1A_2Y_3$ and $f=f_1+f_2$, with the labeling referring to Fig.~\ref{MSSD}. Indeed,  $ \partial_R f = a_{X_1}+a_{X_3}+a_{Y_1}+a_{Y_3} = A_1A_2Y_3\, \partial_R f$. Furthermore, $\tilde{\Phi}_{AAY}(a_{X_3})=1$ and for all other $a \in \{\partial_R f\}$ it holds that $\tilde{\Phi}_{AAY}(a)=0$. Hence, $\tilde{\Phi}(\partial_R f)=1$. The two conditions in Lemma~\ref{L7} are thus satisfied, and the state-dependent version of Mermin's star is contextual.

Let's verify this statement at the elementary level, similar to the symmetry-based contextuality proof for Mermin's square  in Section~\ref{SBMS}. Assume a consistent value assignment exists. Then, with all addition mod 2,
$$
\begin{array}{rcl}
1 &=& s'(a_{XXX})+s'(a_{YXY})\\
&=& s'(a_{X_1})+s'(a_{X_3})+ s'(a_{Y_1})+s'(a_{Y_3})\\
&=& s(a_{Y_1})+\tilde{\Phi}_{AAY}(a_{X_1}) + s(a_{X_3})+\tilde{\Phi}_{AAY}(a_{X_3}) + s(a_{X_1})+\tilde{\Phi}_{AAY}(a_{Y_1}) + s(a_{Y_3})+\tilde{\Phi}_{AAY}(a_{Y_3}) \\
&=& s(a_{X_1})+s(a_{X_3})+ s(a_{Y_1})+s(a_{Y_3}) + 1\\
&=& s(a_{XXX})+s(a_{YXY})+1\\
&=& 0+1+1 =0.
\end{array}
$$ 
Contradiction. Hence no consistent value assignment exists. Above, in lines 1 and 6 we have used that $X_1X_2X_3\,|\text{GHZ}\rangle = -Y_1X_2Y_3\,|\text{GHZ}\rangle = |\text{GHZ}\rangle$. In lines 2 and 5 we have used the consistency of value assignments, in line 3 Lemma~\ref{ConsTrans},
 and in line 4 the above stated values for $\tilde{\Phi}_{AAY}$.
\medskip

In analogy with Lemma~\ref{C1} a cohomological formulation of Lemma~\ref{L7} can be achieved. Let $B_1'$ denote the boundaries in $C_2(E,E_\Psi)$. Let  $U_\Psi$ denote the $1$-cochains defined on the boundaries $B_1'$, and $V_\Psi$ denote $1$-cochains which vanish on the boundaries $B_1'$. With these definitions we have a short exact sequence
\begin{equation}\label{exactchainrel}
0\rightarrow V_\Psi\rightarrow C_1(E,E_\Psi)\rightarrow U_\Psi\rightarrow 0
\end{equation}
a state-dependent version of the short exact sequence (\ref{exatchain}). 

\begin{Lemma}\label{C1SD}
If $[\tilde\Phi|_{B_1'}]\neq 0$ in $H^1(H,U_\Psi)$ then the pair $(\mathcal{O},|\Psi\rangle)$ with symmetry group $H\subset G$ exhibits state-dependent contextuality.
\end{Lemma}
Proof of this lemma is the same as Lemma~\ref{C1} after $\partial$ is replaced by $\partial_R$. 

We are now in the position to obtain the state-dependent versions of Theorems \ref{C1b} and \ref{C1c}. Recall that $N$ is the normal subgroup of $G$ which preserves operators in $\cal{O}$ up to a scalar. Consider the intersection $N'=H\cap N$ and the quotient group $Q'=H/N'$. Let $\Phi':Q'\rightarrow U_\Psi$ denote the composition of a section $\theta':Q'\rightarrow H$ of the  map $H\rightarrow H/N'$ with the restricted map $\tilde\Phi|_{B_1'}:G\rightarrow U_\Psi$. Arguing as in the proof of Theorem \ref{C1b} we obtain the following.

\begin{Theorem}\label{C1bSD}
If $[\Phi']\neq 0$ in $H^1(Q',U_\Psi)$ then the pair $(\mathcal{O},|\Psi\rangle)$ with symmetry group $H\subset G$ exhibits state-dependent contextuality.
\end{Theorem}

Similarly as in Theorem \ref{C1c} second cohomology groups play a role in state-dependent case. The long exact sequence associated to (\ref{exactchainrel}) gives a map $\sigma:H^1(Q',U_\Psi)\rightarrow H^2(Q',V_\Psi)$.

\begin{Theorem}\label{C1cSD}
If $\sigma([\Phi'])\neq 0$ in $H^2(Q',V_\Psi)$ then the pair $(\mathcal{O},|\Psi\rangle)$ with symmetry group $H\subset G$ exhibits state-dependent contextuality.
\end{Theorem}

\section{Conclusion}

In this work we have discussed two kinds of contextuality proofs, based on parity and on symmetry respectively. Both types of proofs come in two flavours, state-independent and state-dependent. For each of the four resulting cases, we have established that the obstruction to the existence of non-contextual hidden variable models is topological. 

Regarding the parity-based proofs (as in Mermin's square and star), algebraic relations among the observables involved are captured by a 2-cocyle $\beta$ living in a suitably defined chain complex ${\cal{C}}_*$, and $[\beta] \not\in H^2({\cal{C}^*},\mathbb{Z}_d)$ is a witness of contextuality.

The symmetry-based proofs invoke transformations that leave the complex ${\cal{C}}_*$ and product relations among commuting observables invariant. Again, nontrivial cohomology of any such group is an obstruction to the viability of a non-contextual hidden variable model for the given setting.

The purpose of studying the above contextuality proofs is their relation to quantum computation. Contextuality has previously been established as a necessary resource for quantum computation, in both the models of  quantum computation with magic states (see \cite{How}--\cite{Qubit}) and measurement-based quantum computation (MBQC) (see \cite{AB}--\cite{RR13}). The type of contextuality considered here is precisely what shows up in MBQC. The study of the mathematical structure underlying such contextuality proofs may thus lead to novel insights into the foundations of quantum computation.

\paragraph{Acknowledgments.} CO acknowledges funding from NSERC.  SDB acknowledges support from the ARC via the Centre of Excellence in Engineered Quantum Systems (EQuS), project number CE110001013. RR is supported by NSERC and Cifar, and  is scholar of the Cifar Quantum Information Processing program.

\appendix

\section{Contextuality in measurement-based quantum computation}\label{ContextMBQC}

In this appendix we review measurement-based quantum computation and the role of contextuality in it. This section is based on \cite{RB01}, \cite{AB}, \cite{Hob} and \cite{RR13}. We assemble this material here is to provide the background and motivation for the cohomological framework of contextuality developed in the main text. 

We emphasize one point in particular: The classical processing relations of MBQC for determining the computational output from the individual measurement outcomes, when spelled out for all values of the computational input, are precisely the equations that give contextuality proofs based on the impossibility of non-contextual value assignments \cite{Merm}. See Section~\ref{star}.

\subsection{Quantum computation by local measurement}\label{MBQC}

Measurement-based quantum computation \cite{RB01} is a scheme of universal quantum computation in which the process of computation is driven by local measurements as opposed to unitary gates. The measurements are applied to a suitable entangled state, such as a cluster state or graph state. The pattern of measurements encodes the algorithm implemented. For reviews of measurement-based quantum computation, see \cite{RBB03}, \cite{CLN}, \cite{RW12}. 

Each MBQC consists of (i) a resource state $|\Phi\rangle$ whose entanglement is consumed by the process of computation, (ii) the set of observables measured to drive the computation, and (iii) rules for the classical side-processing of measurement outcomes.

(i) {\em{Resource state.}} The standard choices for the resource state $|\Phi\rangle$ are cluster states or graph states, which are stabilizer states where the stabilizer generators have a particular geometric interpretation; See \cite{RB01}. 

(ii) {\em{Measured observables.}} The standard choice for the local measured observables is
\begin{equation}\label{Obs}
O_i[q_i] = \cos \phi_i\, X_i + (-1)^{q_i}\sin \phi\, Y_i,
\end{equation}
for all qubits $i$. Therein, the angles $\{\phi_i\}$ are a property of the quantum algorithm to be implemented, and the binary numbers $\{q_i\}$ depend on the classical input to the computation, as well as an offset determined at runtime. The classical input may e.g. be the argument of a function to be evaluated.

(iii) {\em{Classical side-processing.}} The need for classical side-processing in MBQC arises because quantum-mechanical measurement is inherently random. In fact, in the standard scheme \cite{RB01} of MBQC, every individual local measurement is completely random. This has two consequences. First, the classical output is represented by certain {\em{correlations}} of measurement outcomes; only they can be non-random. Second, to keep the computation on track in the presence of randomness, measurement bases need to be adapted according to outcomes obtained in earlier measurements. This boils down to adjusting the parameters $q_i$.

In summary, both the bitwise output $\textbf{o}=(o_1,o_2..,o_k)$ and the choice of measurement bases, $\textbf{q}=(q_1,q_2,..,q_N)$ are functions of the measurement outcomes $\textbf{s}=(s_1,s_2,..,s_N)$. In addition, $\textbf{q}$ is also a function of the classical input $\textbf{i}=(i_1,i_2,..,i_m)$. Remarkably, in standard MBQC these functional relations are all mod 2 linear,
\begin{subequations}\label{CCR}
\begin{align}\label{CCR_out}
\textbf{o}&=Z\textbf{s} \mod 2,\\ 
\label{CCR_in}
\textbf{q} &=T\textbf{s}+S\textbf{i} \mod 2.
\end{align}
\end{subequations}
Therein, the binary matrix $T$ encodes the temporal order in a given MBQC. If $T_{ij}=1$ then the measurement basis at location $i$ depends on the measurement outcome at location $j$, hence the qubit at $j$ must be measured before the qubit at $i$. Therefore, for the measurement events to have a partial (temporal) ordering, the matrix $T$ must be lower triangular w.r.t. a suitable labeling of the qubits.

\subsection{MBQC and Mermin's star}\label{star}

The role of contextuality for measurement-based quantum computation was first noted in the example of Mermin's star \cite{AB}. Here, we review this example.

The state-dependent Mermin star was already discussed in Section~\ref{sdpp}. In the state-dependent version, one of the five contexts of the star is taken up by a quantum state, namely the Greenberger-Horne-Zeilinger  (GHZ) state \cite{GHZ}. The four non-local observables in this context, $X_1X_2X_3$, $X_1Y_2Y_3$, $Y_1X_2Y_3$, $Y_1Y_2X_3$, are stabilizer operators for the GHZ-state. The other four contexts remain for measurement. They are labeled by the elements of the input group $Q=\mathbb{Z}_2 \times \mathbb{Z}_2$; See Fig.~\ref{MSSD2}a.

We now describe the objects (i) - (iii) specifying an MBQC with Mermin's star.
\begin{itemize}
\item[(i)]{The resource state is $|GHZ\rangle= (|000\rangle + |111\rangle)/\sqrt{2}$.}
\item[(ii)]{The local measurable observables are 
\begin{equation}\label{ghzObs}
O_i[0]=X_i, \; O_i[1]=Y_i,\;\; \text{for }i=1,..,3.
\end{equation}}
\item[(iii)]{There are three qubits, two bits of input, $\textbf{i}=(a,b)$, and one bit $o$ of output. The temporal order is flat, $T=0$. The classical side-processing relations Eq.~(\ref{CCR}) are in this case
\begin{subequations}\label{CCRghz}
\begin{align}\label{CCR_outGHZ}
o&= s_1+s_2+s_3 \mod 2,\\ 
\label{CCR_inGHZ}
\left(\begin{array}{c} q_1\\q_2\\q_3\end{array}\right) &=\left(\begin{array}{cc} 1 & 0\\ 0 & 1\\ 1& 1\end{array}\right)\left(\begin{array}{c} a\\ b \end{array}\right) \mod 2.
\end{align}
\end{subequations}}
\end{itemize}
Looping through the possible values for $(a,b)$, with Eqs.~(\ref{CCR_inGHZ}) and (\ref{ghzObs}), Eq.~(\ref{CCR_outGHZ}) becomes four equations, one for each value of $(a,b)$,
\begin{equation}\label{outputEq}
\begin{array}{rcl}
o(0,0) &=& s_{00}(X_1) + s_{00}(X_2) + s_{00}(X_3) \mod 2,\\
o(0,1) &=& s_{01}(X_1) + s_{01}(Y_2) + s_{01}(Y_3) \mod 2,\\
o(1,0) &=& s_{10}(Y_1) + s_{10}(X_2) + s_{10}(Y_3) \mod 2,\\
o(1,1) &=& s_{11}(Y_1) + s_{11}(Y_2) + s_{11}(X_3) \mod 2.
\end{array}
\end{equation}
Therein, $s_{ij}(O)\in \mathbb{Z}_2$ is the outcome of the measurement of an observable $O$ with eigenvalues $\pm1$ only, in the measurement context defined by the input $(i,j)$. 

One may look at Eq.~(\ref{outputEq}) from the quantum mechanical and the HVM angle, which we will do in turn. The GHZ-state satisfies the eigenvalue equations
$$
X_1X_2X_3 |GHZ\rangle = -X_1Y_2Y_3 |GHZ\rangle = -Y_1X_2Y_3 |GHZ\rangle = -Y_1Y_2X_3|GHZ\rangle = |GHZ\rangle.
$$
Further, since  the observables $X_1$, $X_2$ and $X_3$ pairwise commute and obey the relation $X_1X_2X_3 = (X_1I_2I_3)(I_1X_2I_3)(I_1I_2X_3)$, it holds that $s_{00}(X_1) + s_{00}(X_2)+s_{00}(X_3) \mod 2=s_{00}(X_1X_2X_3)$. With the first of the above eigenvalue equations, $s_{00}(X_1X_2X_3)=0$, we thus have $o(0,0)=0$ with certainty. By the same argument, $o(0,1)=o(1,0)=o(1,1)=1$. Thus, the quantum mechanical prediction is that the computation described evaluates the function
\begin{equation}\label{OR}
o(a,b) = a\, \text{OR} \, b.
\end{equation}
This is of significance from the following fundamental point of view. The classical control computer of MBQC by itself is only capable of performing mod 2 addition, cf. Eq.~(\ref{CCR}). Hence it is not classically universal. If supplemented with quantum resources---GHZ states and the capability to measure local Pauli observables $X_i$, $Y_i$---it can execute OR-gates in addition, and thereby becomes classically universal. The computational power of the control computer is thus significantly boosted.

Let's now look at Eq.~(\ref{outputEq}) from the perspective of a non-contextual HVM with deterministic value assignments.  Non-contextual HVMs with definite value assignments invoke  assumption the additional assumption that the ``pre-existing'' values of measurement outcomes are independent of the measurement context,
\begin{equation}
\label{ncHVMass}
s_{ij}(O) =s(O),\; \forall O \in \Omega,\, \forall (i,j) \in \mathbb{Z}_2\times \mathbb{Z}_2.
\end{equation}

Can there be a consistent non-contextual assignment of values $s(X_1), .., s(Y_3)$ on the r.h.s. of Eq.~(\ref{outputEq})?---This is quickly ruled out. Substituting Eq.~(\ref{ncHVMass}) into Eq.~(\ref{outputEq}) and adding the four resulting equations mod 2 leads to the familiar contradiction $1=0$. Hence a consistent non-contextual HVM value assignment does not exist.\smallskip

We observe that the above statements about computational power and contextuality do not require the function $o$ to be precisely an OR-gate. The classical control computer is boosted to classical universality whenever the function $o$ is {\em{non-linear}}, i.e., if and only if $\Sigma(o):=o(0,0)+o(0,1)+o(1,0)+o(1,1)\mod 2 =1$. The same relation is an obstruction to the existence of an ncHVM. To summarize, $\Sigma(o) =1$ is both a witness of contextuality and a guarantee for boosting the a priori very limited classical control computer to classical universality.

\subsection{Computational output and contextuality}

The points made in the last paragraph about the MBQC based on Mermin's star generalize to all MBQCs that satisfy the classical processing relations Eq.~(\ref{CCR}); See \cite{Hob}, \cite{RR13}.   When Eq.~(\ref{CCR_out}), which defines the MBQC output, is spelled out for all input values and combined with the ncHVM assumption Eq.~(\ref{ncHVMass}), those very equations rule out the existence of a corresponding non-contextual HVM. Furthermore, it is the non-linearity of the outputted function (and hence the boost in classical computational power) that represents the obstruction to the existence of non-contextual HVMs.

Our running example of the MBQC based on Mermin's star misses two aspects of the general case. First, it is temporally flat, i.e., measurement bases are not influenced by the outcomes  of measurements on other qubits, and second, it is deterministic. Both of these constraints  can be relaxed while keeping the relation with contextuality. We have the following result.
\begin{Theorem}{\em{\cite{RR13}}}
Be ${\cal{M}}$ an MBQC with classical processing relations Eq.~(\ref{CCR}) evaluating a function $o:(\mathbb{Z}_2)^m \longrightarrow \mathbb{Z}_2$. Then, ${\cal{M}}$ is contextual if it succeeds with an average probability $p_S>1-d_H(o)/2^m$, where $d_H(o)$ is the Hamming distance of $o$ from the closest linear function.
\end{Theorem}

{\em{Remark:}} The lowest contextuality thresholds are reached for bent functions. For $m$ even and $o$ bent, it holds that $d_H(o) = 2^{m-1} - 2^{m/2-1}$ \cite{MWS}, and therefore the contextuality threshold for the average success probability $p_S$ approaches $1/2$ for large $m$. An MBQC can thus be contextual even if its output is very close to completely random.

\section{Chain complexes}

Throughout the text we work with modules over the ring $\mathbb{Z}_d=\{0,1,\cdots,d-1 \}$.
A chain complex of modules is a sequence 
$$
{\cal C}_*: \cdots \rightarrow C_n \stackrel{\partial_n}{\rightarrow} C_{n-1} \stackrel{\partial_{n-1}}{\rightarrow} C_{n-2} \cdots 
$$
such that the composition of any two successive maps gives zero i.e. $\partial\partial=0$. Homology groups of the chain complex are defined by
$$
H_n({\cal C}_*)=\frac{{\text{ker}} (\partial_{n})}{{\text{im}} (\partial_{n+1})}.
$$
A map $f:{\cal C}\rightarrow {\cal D}$ of chain complexes is a sequence of module maps $C_n\rightarrow D_n$ which commutes with the differential $\partial$. Such a map induces a map in homology $f_*:H_n({\cal C}_*)\rightarrow H_n({\cal D})$.

Dually, we can  consider a cochain complex obtained from a chain complex. This is a sequence
$$
{\cal C}^*: \cdots \rightarrow C^{n-2} \stackrel{d_{n-2}}{\rightarrow} C^{n-1} \stackrel{d_{n-1}}{\rightarrow} C^{n} \cdots 
$$
where $C^n$ consists of module maps  $\alpha:C_n\rightarrow \mathbb{Z}_d$, and $d_n$ is defined by $d_n(\alpha)(c) = \alpha(\partial_{n+1}(c))$ for all $c$ in $C_{n+1}$. Similarly we can talk about cohomology groups
$$
H^n({\cal C}^*)=\frac{{\text{ker}} (d_{n})}{{\text{im}} (d_{n-1})}.
$$
A map $f$ of chain complexes as above induces a map in cohomology $f^*:H^n({\cal D}^*)\rightarrow H^n({\cal C}^*)$ in the reverse direction. 

A simplicial complex with edges, faces, and volumes... naturally defines a chain complex. The modules $C_0$, $C_1$, $C_2$, $C_3$... in this complex consists of $\mathbb{Z}_d$-linear combinations of labels representing vertices, edges, faces, volumes... Another source for a chain complex is group cohomology. Starting from a single vertex, one builds a space by glueing the boundary of an edge  representing an element $g\in G$. The resulting space is a bouquet of circles where the circles are labelled by the elements of the group. Now continue to glue higher dimensional basic shapes which encode the structure of the group. For each pair of group elements $(g_1,g_2)$ glue a triangle whose edges are $g_1$, $g_2$, and $g_1g_2$. This process repeats for higher dimensional triangles which corresponds to an $n$-tuple $(g_1,g_2,\cdots,g_n)$ of group elements so that edges are products of these elements arranged in an organized way. The resulting space is called the classifying space of $G$. The associated chain complex in dimension $n$ is a module which consists of $\mathbb{Z}_d$-linear combinations of the representatives $[g_1|g_2|\cdots|g_n]$. The cochain complex consists of  $\mathbb{Z}_d$-linear combinations of set maps $G^n \rightarrow \mathbb{Z}_d$. It is a standard fact in group cohomology that it suffices to consider non-trivial $n$-tuples i.e. $g_i\not=1$ for all $i$. This is convenient for computational purposes.  

In the text we introduce a complex ${\cal C}_*(E)$ constructed from commuting operators which imitates the construction of a classifying space.
We will show that why ${\cal C}_*(E)$ is a chain complex i.e. $\partial \partial =0$.  The proof is similar to  the group cohomology case. Let $[a_1|\cdots|a_n]$ be a basis element of $C_n(E)$. The $n$-tuple consists of commuting elements in $E$. Although our complex consists of dimensions $n=0,1,2,3$ we prove the result for all $n$.
Let us introduce the following maps
$d_0[a_1|\cdots|a_n]=[a_2|\cdots|a_n]$, $d_i[a_1|\cdots|a_n]=[a_1|\cdots|a_i+a_{i+1}|\cdots|a_n]$ for $1\leq i\leq n-1$, and $d_n[a_1|\cdots|a_n]=[a_1|\cdots|a_{n-1}]$. Then we can write $\partial =\sum_{i=0}^n (-1)^n d_i$. 
As a preliminary observation one checks that by definition  $d_id_j = d_{j-1}d_i$ for $i\leq j-1$. Using this
\begin{eqnarray*}
\partial \partial &=& \sum_{i=0}^{n-1} \sum_{j=0}^n (-1)^{i+j} d_id_j\\
 &=& \sum_{i\leq j-1} (-1)^{i+j} d_id_j + \sum_{i\geq j} (-1)^{i+j} d_id_j\\
 &=& \sum_{i\leq j-1} (-1)^{i+j} d_{j-1}d_i + \sum_{i\geq j} (-1)^{i+j} d_id_j \\
  &=& \sum_{i\leq k} (-1)^{i+k+1} d_{k}d_i + \sum_{j\leq i} (-1)^{i+j} d_id_j=0.
\end{eqnarray*}
In the last sum we set $k=j-1$ hence $0\leq k\leq n-1$. The first sum is indexed over $\{ 0\leq i \leq n-1  \text{ and } 0\leq k\leq n-1|\;i\leq k\}$ and the second one is indexed over $\{ 0\leq i\leq n-1 \text{ and } 0\leq j \leq n |\; i\geq j\}$. Note that these sets are the same. Therefore two sums cancel each other when corresponding terms with different signs are matched together\footnote{As an example consider $\partial\partial:C_2\rightarrow C_0$. In this case the first sum is $-d_0d_1 +d_0d_2-d_1d_2$ and the second sum is $d_0d_0 -d_1d_0+d_1d_1$. }. 
 
\section{A converse of Lemma~\ref{split}}\label{C}

Lemma~\ref{split} has a converse if the symmetry group $G$ is large enough. We start with an observation which will lead to a structural relation between the symmetry group and the chain complex.  Recall the definition of the sub-complex $V=\{\alpha\in C^1|\; d^v\alpha=0\}$. In particular, this is an abelian group under addition. We  define an action of $V$ on the set of operators by
$$
\alpha(T_a)=\omega^{\alpha(a)}T_a
$$
for each $\alpha\in V$. Note that this gives a group action since $(\alpha+\alpha')(T_a)=\omega^{\alpha(a)+\alpha'(a)}T_a= \alpha(\alpha'(T_a))$. It also satisfies Eq.~(\ref{TTrans2}). Therefore this is a symmetry of the system. We can regard this symmetry as a group homomorphism
$$
i:V\rightarrow \Aut({\cal O})
$$
which is in fact injective. We  identify  $V$ as a subgroup of $\Aut({\cal O})$.  

In general given a symmetry group $G$ we defined $N$ as the subgroup which fixes each edge:
$$
n(T_a)=\omega^{\tilde\Phi_n(a)}T_a.
$$ 
Given a symmetry associated to the homomorphism $\xi:G\rightarrow \Aut({\cal O})$  in Eq.~(\ref{rep}) the image $\xi(N)$ of $N$ lies inside $V$. That is the restriction of $\xi$ to $N$ gives a  group homomorphism
$$
\xi|_N:N\rightarrow V \subset \Aut({\cal O})
$$
which sends $n$ to $\tilde\Phi_n$.

\begin{Lemma}\label{converse}
Assume that $\xi:G\rightarrow\Aut({\cal O})$ is injective and $\xi|_N:N\rightarrow V$ is an isomorphism. Then $G$ splits as $Q\ltimes N$ if and only if $\sigma([\Phi])=0$.

\end{Lemma}
{\em{Proof of Lemma~\ref{converse}.}} If $G$ splits then $\sigma([\Phi])=0$ is proved in Lemma~\ref{split}. For the converse assume that $\sigma([\Phi])=0$ that is  there exists a $\chi: Q \longrightarrow V$ such that 
\begin{equation}\label{cohoTr}
d^h\Phi'+d^h\chi =0.
\end{equation}
Next, corresponding to the map $\theta: Q\longrightarrow G$ we define a new map $\hat\theta$ via $\hat\theta(q) = \theta(q)n_q$, where $n_q\in N$ is such that $n_q(T_a) =\omega^{\chi_q(a)}T_a$, for all $q\in Q$ and all $a \in E$. Under the assumption that $N\cong V$, such an $n_q$ exists for all $q\in Q$.

Now, the action of $\hat\theta(Q)$ on the $T_a$ is 
$$
(\hat\theta(q))(T_a) = (\theta(q)n_q)(T_a) =  \theta(q)(n_q(T_a)) = \omega^{\chi_q(a)+\Phi'_q(a)}T_{\theta(q)(a)} =   \omega^{\Phi''_q(a)}T_{\theta(q)(a)},
$$
where $\Phi'':=\Phi'+\chi$. We can now show that $\hat\theta(pq) = \hat\theta(p)\hat\theta(q)$, namely
$$
\begin{array}{rcl}
(\hat\theta(p)\hat\theta(q))(T_a) &=& \theta(p)(\theta(q)(T_a))\\
&=& \omega^{\Phi''_q(a)+\Phi''_p(\theta(q)a)}T_{\theta(p)\theta(q)(a)}\\
&=& \omega^{\Phi''_{pq}q(a)}T_{\theta(p)\theta(q)(a)}\\
&=& \hat\theta(pq)(T_a).
\end{array}
$$ 
Therein, in the third line we have used  Eq.~(\ref{cohoTr}). Thus $G = Q\ltimes N$. $\Box$

\end{document}